\documentclass[12pt,fleqn]{article}
\usepackage{verbatim,amssymb,amsmath,graphics,amsthm,cite}
\numberwithin{equation}{section}

\theoremstyle{definition}

\newtheorem{proposition}{Proposition}[section]
\newtheorem{lemma}{Lemma}[section]

\theoremstyle{definition}
\newtheorem{definition}{Definition}[section]

\renewcommand{\l}{\Lambda}
\renewcommand{\H}{{\cal H}}
\renewcommand{\P}{{\cal P}}
\newcommand{\bs}{\boldsymbol}
\newcommand{\p}{\hat{p}}
\newcommand{\<}{\langle}
\newcommand{\be}{\begin{equation}}
\renewcommand{\>}{\rangle}

\newcommand{\sm}{\mathsf{s}}

\renewcommand{\a}{\bs{\p}j_{3}[\sm j]}

\newcommand{\am}{\bs{\p}j_{3}[\sm j]^{-}}
\newcommand{\ap}{\bs{\p}j_{3}[\sm j]^{+}}
\newcommand{\amp}{\bs{\p}j_{3}[\sm j]^{\mp}}
\newcommand{\amd}{\bs{\p}j_{3}[\sm_{R} j_{R}]^{-}}
\newcommand{\amr}{\bs{\p}j_{3}[\sm j_{R}]^{-}}

\newcommand{\bp}{\tilde{\cal S}\cap{\cal H}_{+}^{2}|_{{\mathbb R}_{\sm_0}}}
\newcommand{\bm}{\tilde{\cal S}\cap{\cal H}_{-}^{2}|_{{\mathbb R}_{\sm_0}}}
\newcommand{\Bp}{\tilde{\cal S}\cap{\cal H}_{+}^{2}}
\newcommand{\Bm}{\tilde{\cal S}\cap{\cal H}_{-}^{2}}

\newcommand{\Bmp}{\tilde{\cal S}\cap{\cal H}_{\mp}^{2}}
\newcommand{\Bpm}{\tilde{\cal S}\cap{\cal H}_{\pm}^{2}}

\begin{document}
\title{Time Asymmetric Quantum Theory\\
III. Decaying States and the Causal Poincar\'e Semigroup.}
\author{A.~Bohm\footnote{bohm@physics.utexas.edu} 
\qquad H.~Kaldass\footnote{Present address: Deutsches Elektronen-Synchrotron, DESY, Platanenallee 6, D-15738 Zeuthen, e.mail hani@ifh.de}\qquad S.~Wickramasekara\footnote{sujeewa@physics.utexas.edu} 
\\ Physics Department\\ The University of Texas at Austin\\
Austin, Texas 78712} 
\maketitle
\begin{abstract}
A relativistic resonance which was defined by a pole of the \mbox{$S$-matrix},
or by a relativistic Breit-Wigner line shape, is
represented by a generalized state vector (ket) which can be obtained
by analytic extension of the relativistic Lippmann-Schwinger kets. 
These Gamow kets span an irreducible representation space for
Poincar\'e transformations which, similar to the Wigner
representations for stable particles, are characterized by spin
(angular momentum of the partial wave amplitude) and complex mass (position
of the resonance pole). The Poincar\'e transformations of the Gamow
kets, as well as of the Lippmann-Schwinger plane wave scattering
states, form only a semigroup of Poincar\'e
transformations into the forward light cone. Their transformation
properties are derived. From these one obtains an unambiguous definition
of resonance mass and width for relativistic resonances.
The physical interpretation of these transformations for
the Born probabilities and the problem of
causality in relativistic quantum physics is discussed.
\end{abstract}
\maketitle

\section{Introduction}\label{summary}
In this paper, we investigate the properties
of the relativistic Gamow vectors under Poincar\'e transformations.  
The relativistic Gamow vectors were defined in \cite{paper1}.
They provide a state vector description for unstable particles.
An unstable particle is usually
associated with the pole of the relativistic $S$-matrix
element with angular momentum $j_R$ (spin of the resonance)
at the complex invariant mass squared $\sm=\sm_R$. In order to obtain the 
$j_R$-th partial $S$-matrix element, we have to use basis
vectors of total angular momentum
for the out-states of the decay products. These angular
momentum basis vectors are obtained when the 
space of decay products is resolved into a continuous direct
sum of irreducible representation (irrep) spaces of the Poincar\'e
group ${\P}$ \cite{wightman, joos, macfarlane}. In the case
where the (asymptotically free) decay products consist
of two particles, with each one furnishing a unitary irreducible
representation (UIR) space of ${\P}$ labeled by the mass $m_{i}$ and 
the spin $s_{i}$,
$\H(m_i,s_i)$, $i=1,\,\,2$, the direct product space of the two-particle
system $\H_{12}=\H(m_1,s_1)\otimes\H(m_2,s_2)$ is reduced into a 
direct sum of UIR spaces \cite{wightman, joos, macfarlane,hani}according to
\begin{equation}
\label{star}
\H_{12}=\sum_{j\eta}\int_{(m_1+m_2)^{2}}^{\infty}d\mu(\sm)
\oplus\H^{\eta}_{n}(\sm,j)\,.
\end{equation}
In \eqref{star}, $\eta$ and $n$ are degeneracy and particle species labels, 
respectively, and
$\sm$ is the invariant mass squared
for the two-particle system, $\sm=p^2=(p_1+p_2)^{2}$.
In place of the usual momentum eigenkets of the Wigner basis we
use in \eqref{star} the eigenkets of
4-velocity $|\a\eta,n\>$, $\p=p/\sqrt{\sm}$.
The velocity eigenkets $|\a\eta,n\>$ furnish, like the momentum kets,
a complete system of basis vectors of \eqref{star} \cite{hani}.
This means every vector $\phi$
can be written as the continuous linear superposition
\begin{equation}
\label{diracexpansion}
\phi=\sum_{jj_{3}\eta}\int_{(m_1+m_2)^{2}}^{\infty}d\sm
\int\frac{d^{3}\p}{2\p^0}|\a\eta,n\>\<\a\eta,n|\phi\>\,.
\end{equation}
The statement \eqref{diracexpansion}
is Dirac's basis vector expansion 
for a complete set of commuting observables (self-adjoint operators)
and is one of the basic rules used in quantum theory.
Its mathematical justification is given by
the Nuclear Spectral Theorem proved for a 
Rigged Hilbert Space $\Phi\subset\H\subset\Phi^\times$ and 
\eqref{diracexpansion} holds for every $\phi\in\Phi$~\cite{gelfand}. 
The kets $|\a\eta,n\>$ are generalized eigenvectors 
of the mass operator $M=(P_\mu
P^\mu)^{1/2}=[(P_1+P_2)_{\mu}(P_1+P_2)^{\mu}]^{1/2}$ 
with eigenvalue $\sqrt{\sm}$ and of the $4$-velocity
operators $\hat{P}\equiv M^{-1}(P_1+P_2)$ with eigenvalue $\p$.
How these kets can be constructed in terms of the $4$-velocity eigenkets
of the product space $\H(m_i,s_i)$ using the Clebsh-Gordan 
coefficients has been shown in~\cite{hani}.

The kets $|\a\eta,n\>$ are elements of the dual space $\Phi^\times$, which 
is defined as the space of
$\tau_\Phi$-continuous antilinear functionals on $\Phi$;
and the space $\Phi$ is a dense subspace of ${\cal H}$.
For the space $\Phi$ one
can choose different dense subspaces of the Hilbert space $\H$ 
as long as they fulfill the conditions
for the Nuclear Spectral Theorem~\footnote{E.g., $\Phi$ could be
chosen \cite{istanbul} to be
the subspace of differentiable vectors of a unitary representation of
$\P$ equipped with a 
nuclear topology $\tau_{\Phi}$ defined by the countable number
of norms: $\|\phi\|_{p}=\sqrt{(\phi,(\Delta+1)^{p}\phi)}$, where
$\Delta=\sum_{\mu}\hat{P}_{\mu}^{2}+\sum_{\mu\nu}\frac{1}{2}J_{\mu\nu}^{2}$
is the Nelson operator~\cite{nelson}. But it could also be chosen
differently, which we will do later when we will choose two different
subspaces called $\Phi_-$ and $\Phi_+$ below.}\label{f1}
and thus obtain different rigged Hilbert
spaces $\Phi\subset\H\subset\Phi^\times$ for the same $\H$.

If one starts with the spaces $\H(m_i,s_i)$ of asymptotically free
decay products one obtains by reduction into the irreducible
representation $\H^\eta_n(\sm,j)$ 
the four-velocity basis vectors $|\a\eta,n\>$ of the interaction-free
Poincar\'e group.  This means the $|\a\eta,n\>$ for any given value of
$\sm$ from the continuous spectrum $(m_1+m_2)^2\leq\sm<\infty$ transform
like kets of the unitary group representation $[\sm,j]$ of Wigner,
cf.~\eqref{v25} below.

The interacting out- and in-state vectors 
$|\a\eta,n^\mp\>$ are obtained by the standard postulate of the existence
of Moeller wave operators $\Omega^\mp$\cite{weinberg}: 
\begin{equation}
\label{moeller}
|\,\bs{\p}j_{3}[{\sm}j]\eta,n^{\mp}\,\rangle
=\Omega^{\mp}|\,\bs{\p}j_{3}[{\sm}j]\eta,n\,\rangle\,.
\end{equation}
This means that the interaction eigenkets 
$|\,\bs{\p}j_{3}[{\sm}j]\eta,n^{\mp}\,\rangle$
are assumed to be connected to the interaction-free kets 
$|\,\bs{\p}j_{3}[{\sm}j]\eta,n\,\rangle$ by the Lippmann-Schwinger
equation. In Section~$3$ of \cite{paper1} it was explained that a 
backdrop to an interaction free asymptotic theory with interaction-free in- and
out-states and interaction-free basis vectors as postulated by \eqref{moeller} is not
needed. The theory can be formulated in terms of interaction-incorporating state-/observable-vectors,
$\phi^+/\psi^-$ defined mathematically by the Hardy spaces $\Phi_-$/$\Phi_+$ and
defined physically by the preparation/registration apparatuses in the asymptotic
region, and in terms of the basis kets $|\bs{\p}j_3[\sm j]\eta,n^\mp\>\in\Phi_\pm^\times$ upon
which the interaction-incorporating ``exact generators'' \cite{weinberg} of 
the Poincar\'e transformations act. The in- and out-boundary conditions formulated usually
by the infinitesimal imaginary part of $\sm$ (or equivalently of $p^0$~\footnote{cf.~footnote~$15$ of
\cite{paper1}}) of the Lippmann-Schwinger equation, is now contained in the Hardy space
postulate for the space of in-states $\Phi_-$ and for the space of out-observables $\Phi_+$,
cf.~\eqref{pr} below.
As we shall explain in Section \ref{semigroup}, the kets 
$|\bs{\p}j_3[\sm j]\eta n^\pm\>$ 
do not furnish anymore a 
representation of the whole Poincar\'e group; this  could have been 
suspected already from the infinitesimal
imaginary part of $\sm$ in the Lippmann-Schwinger equations.

For notational convenience, we will drop the labels $\eta$, $n$
(also for the case $s_1=s_2=0$ the degeneracy labels are 
not needed~\cite{joos,macfarlane,hani})
and denote the out/in two-particle basis vectors by $|\amp\>$.
Usually, they are defined as (generalized) 
eigenvectors of the mass operator $M=(P_\mu P^\mu)^{1/2}$,
and of the $4$-momentum operators of the Poincar\'e group that 
incorporates interaction~\cite{weinberg}. We will use in place 
of the usual  momentum kets, the eigenvectors of the 
$4$-velocity operator $\hat{\bs{P}}={\bs{P}}M^{-1}$ for reasons explained 
in~\cite{paper1}.
Their eigenvalues are:\footnote{For the notion of generalized eigenvectors, as well
as for the definition of conjugate operators 
($M^\times$, $P^{\mu\times}$ in \eqref{mtimes}
and \eqref{2.15.5}) see Appendix~A in \cite{paper1}.}
\begin{equation}
\label{mtimes}
\begin{split}
M^{\times}|\amp\>=\sqrt{\sm}|\amp\>\quad(m_1+m_2)^{2}\equiv m_0^{2}\leq \sm<
\infty\\
\hat{P}^{\mu\,\times}|\amp\>=\p^\mu|\amp\>\quad{\bs \p}\in{\mathbb R}^{3}\,;
\quad \p^0=\sqrt{1+\bs{\p}^{2}}\,.
\end{split}
\end{equation}
The eigenvalues of the exact Hamiltonian $H$ and of the exact momentum
operators are
\begin{equation}
\label{2.15.5}
\begin{split}
H^\times|\amp\>&=\gamma\sqrt{\sm}|\amp\>\\
\bs{P}^\times|\amp\>&=\gamma\sqrt{\sm}\bs{v}|\amp\>\,,
\end{split}
\end{equation}
where $\bs{v}$ is the three-velocity, 
\begin{equation}
\label{1.5a}
\tag{\ref{2.15.5}a}
\bs{\p}=\gamma\bs{v},\ {\rm and}\  
\gamma=1/\sqrt{1-\bs{v}^{2}}=\sqrt{1+\bs{\p}^{2}}=\p_{0}
\end{equation}

Like the free kets $|\a\>$, 
the interacting out/in kets $|\amp\>$ are also basis vectors 
of Rigged Hilbert Spaces. We choose for the out-kets and for the
in-kets {\em two}
different Rigged Hilbert Spaces \cite{paper1} 
\begin{equation}
\label{2rhs}
\Phi_\pm\subset\H\subset\Phi_\pm^\times
\end{equation}
with the same Hilbert space $\H$. 
This choice is suggested by
the different properties for the 
two Lippmann-Schwinger equations with $\mp i0$.
In particular we denote the in-states by $\phi^+$ and the
space of in-states by $\Phi_-$; then for every $\phi^+\in\Phi_-$
we have the Dirac basis vector expansion (nuclear spectral theorem for
the RHS \eqref{2rhs}) 
\begin{equation}
\label{t1}
\phi^+=\sum_{jj_{3}}\int_{(m_1+m_2)^{2}}^{\infty}d\sm
\int\frac{d^{3}\p}{2\p^0}|\ap\>\<^{+}\a|\phi^+\>\,.
\end{equation}
And we denote the out-states by $\psi^-$ and the space of out-states
by $\Phi_+$; then for every $\psi^-\in\Phi_+$,
\begin{equation}
\label{t2}
\psi^-=\sum_{jj_3}\int_{(m_1+m_2)^2}^\infty d\sm\int 
\frac{d^{3}\p}{2\p^0}|\am\>\<^{-}\a|\psi^-\>\,.
\end{equation}
Though the Dirac kets $|\amp\>$ are now mathematically defined as 
elements of the space
of continuous antilinear functionals on $\Phi_\pm$, 
$|\amp\>\in\Phi_\pm^\times$, which fulfill the eigenvalue equations
\eqref{mtimes}, \eqref{2.15.5} and not by a Lippmann-Schwinger equation,
we still shall refer to them as Lippmann-Schwinger kets.
The expansions \eqref{t1} and \eqref{t2} are then standard expansions
used in scattering theory, only that usually one does not distinguish
between the spaces $\Phi_+$ and $\Phi_-$ but just talks of the
``same'' Hilbert space \cite{weinberg}, though the kets lie outside
the Hilbert space. 

The choice of the two rigged Hilbert spaces
\eqref{2rhs} means that for the interacting two-particle system of
a resonance scattering experiment we use the following new {\it hypothesis}
of relativistic time asymmetric quantum theory:
There is a Rigged Hilbert Space:
\begin{subequations}
\label{pr}
\begin{equation}
\label{pr1}
\tag{\ref{pr}$_{-}$}
\Phi_-\subset\H\subset\Phi_-^\times\quad
\parbox[t]{3.5in}{for the prepared in-states defined by the
preparation apparatus (accelerator).}
\end{equation}
And there is a different Rigged Hilbert Space with the same
Hilbert space $\H$:
\begin{equation}
\label{pr2}
\tag{\ref{pr}$_{+}$}
\Phi_+\subset\H\subset\Phi_+^\times\quad
\parbox[t]{3.5in}{for the detected out-states or observables defined by the
registration apparatus (detector).}
\end{equation}
\end{subequations}
We thus distinguish meticulously between states $\phi^+\in\Phi_-$ and 
observables $\psi^-\in\Phi_+$, not only in their physical
interpretation but also by their mathematical representation.

Mathematically the space $\Phi_+$ is the nuclear Fr\'echet space
which is realized\footnote{This means the abstract Hardy space of the
upper/lower half-plane $\Phi_\pm$ is defined as the function space in which
the energy wavefunctions $\psi^-(\sm)$/$\psi^+(\sm)$ are well
behaved Hardy functions $\Bpm|_{\mathbb{R}_{\sm_0}}$ analytic  
in the upper/lower half pane, where the label $\pm$ refers in the  
standard notation of mathematics to the upper/lower half plane. The
labels $-/+$ of $\psi^-/\phi^+$ is the most common physics notation
for out/in state vectors. These standard notations of mathematics
and physics we supplemented with a notation that distinguishes between states
$\phi$ and observables $\psi$ since what one calls out-states in scattering 
theory is in most cases an observable defined by a detector.}
by the Hardy functions in the upper half-plane
of the second sheet of the Riemann $\sm$-surface, $\Bp$. Precisely  
\begin{subequations}
\label{t}
\begin{equation}
\label{t3}
\tag{\ref{t}$_+$}
\psi^-\in\Phi_+\text{ if and only if  }
\psi^-(\sm)\equiv\<^{-}\a|\psi^-\>\in\Bp|_{\mathbb{R}_{\sm_0}}
\otimes{\cal S}(\mathbb{R}^3)\,.
\end{equation}
The \addtocounter{footnote}{-1} space $\Phi_-$, in contrast, 
is defined mathematically as
the space which is realized\footnotemark\ by the Hardy functions in the 
lower half-plane of the second sheet of the Riemann $\sm$-surface,
$\Bm$. Precisely\footnote{Note that from \eqref{t3} follows that
$\<\psi^-|\am\>=\<^{-}\a|\psi^-\>^{*}\in\Bm|_{\mathbb{R}_{\sm_0}}\otimes
{\cal S}({\mathbb R}^{3})$ because the complex conjugate of a 
Hardy function from above is a Hardy function from below, and vice
versa. Consequently, the function of $\sm$ in the $S$-matrix element
$(\psi^-,\phi^+)$ -cf $(5.10)$ of \cite{paper1}- can be analytically
continued into the lower complex $\sm$-plane in the second sheet.}
\begin{equation}
\label{t4}
\tag{\ref{t}$_-$}
\phi^+\in\Phi_-\text{ if and only if  }
\phi^+(\sm)\equiv\<^{+}\a|\phi^+\>\in\Bm|_{\mathbb{R}_{\sm_0}}
\otimes{\cal S}(\mathbb{R}^{3})
\end{equation}
\end{subequations}
In \eqref{t}, $\tilde{\cal S}({\mathbb R})$ is a closed subspace of
the Schwartz space ${\cal S}(\mathbb R)$ 
(see Definition~\ref{stilda} of the mathematical
Appendix), 
$\H_{\pm}^{2}$ are the spaces of Hardy class functions from above
and below respectively (see Definition~\ref{h:1} of the mathematical
Appendix) and  
$|_{{\mathbb R}_{\sm_{0}}}$ means restrictions of $\sm$ to the
``physical'' values 
${\mathbb R}_{\sm_{0}}=\{\sm:\ \sm_0=(m_1+m_2)^{2}\leq\sm<\infty\}$.
The spaces $\Bmp$ are for the $\sm$-variable while ${\cal S}({\mathbb R}^{3})$
is for the $\bs{\p}$-variable. The space $\tilde{\cal S}$ was chosen for the
realization of the spaces $\Phi_\mp$ because then $\Phi_-$ and
$\Phi_+$ remain invariant with respect to the action of the generators
of the Poincar\'e group. The Hardy spaces $\H_\mp^{2}$ have been chosen
because --due to the Paley-Wiener theorem \cite{hardy}-- 
they allow a mathematical representation of causality in the 
following way~\cite{antoniou}: A state $\phi^+\in\Phi_-$ must be
prepared first before an observable $\psi^-\in\Phi_+$ can be 
observed in this state.

The hypothesis \eqref{pr},\eqref{t} is the 
new postulate by which our time-asymmetric quantum theory differs
from the postulates of orthodox (von Neumann) 
quantum mechanics which uses the Hilbert space axiom:\\
 
set of in-states $\{\phi^+\}\equiv$ set of 
out-observables~$\{\psi^-\}=\H$,\\
\indent\qquad\qquad\qquad\qquad 
$\<\bs{\p}j_3[\sm j]|\phi\>\in L^2({\mathbb{R}_{\sm_0}},d\sm)\otimes
L^2\left({\mathbb{R}}^3,\frac{d^3\p}{2\p^0}\right)$\\

\noindent or in a milder form, the assumption 
that $\{\phi^+\}\equiv\{\psi^-\}=\Phi$, where $\Phi$ is the dense
subspace in footnote \ref{f1} of $\H$. That means instead of using for
the wave functions $\<^-\hat{\bs{p}},j_3,[\sm,j]|\psi^-\>$ and 
$\<^+\hat{\bs{p}},j_3,[\sm,j]|\phi^+\>$ the same space of smooth
functions of $\sm$, we postulate that these wave functions are smooth
and can also be
analytically continued into the upper and lower half complex
$\sm$-plane, respectively.

In scattering theory one uses already a rudimentary form
of the time asymmetric boundary condition \eqref{pr}, \eqref{t} 
by requiring
that the eigenkets \eqref{moeller} fulfill different Lippmann-Schwinger
equations with $-i0$ or $+i0$ in the denominator.
This implies that $\<^+\bs{\p}j_3[\sm,j]|\phi^+\>$ and
$\<\psi^-|\bs{\p}j_3[\sm,j]^-\>$ must be analytic in at least 
a strip below the 
real $\sm$-axis; we generalize this by requiring in \eqref{t} that they
are analytic and Hardy in the whole lower semiplane of the second
sheet$^4$. Generalized eigenvectors of
\eqref{2.15.5} which are either elements of $\Phi_+^\times$ or of
$\Phi_-^\times$ will therefore be called Dirac-Lippmann-Schwinger
(D-L-Sch) kets, 
in order to distinguish them from the ordinary Dirac kets of 
\eqref{diracexpansion} 
which are usually defined as functionals on the Schwartz space 
(they fulfill time symmetric boundary conditions$^1$). In addition
to the D-L-Sch kets, the spaces $\Phi_\pm^\times$ also
contain other vectors, e.g., 
the Gamow vectors, $(5.28)$ of \cite{paper1}, whose complex 
energy value $\sqrt{\sm}=(M_R-i\Gamma/2)$ has a finite imaginary part. 

The postulate \eqref{pr}~\eqref{t} is the
only new condition we introduce.
All other fundamental postulates of quantum mechanics
remain the same as in conventional
quantum theory (Dirac formulation).

The two Rigged Hilbert Spaces of states~\eqref{pr1} and 
observables~\eqref{pr2} are thus realized (i.e., their space
of wavefunctions are given) by the pair of Rigged Hilbert Spaces
\begin{subequations}
\label{striple}
\begin{equation}
\label{striplet}
\tag{\ref{striple}$_{\mp}$}
\!\!\!\!\!\!\!\!\!\!
\Bmp|_{{\mathbb R}_{{\sm}_0}}\otimes{\cal S}({\mathbb R}^{3})
\subset 
L^{2}({\mathbb R}_{{\mathsf s}_0},d\sm)\otimes
L^{2}({\mathbb R}^{3},\frac{d^{3}\p}{2\p^0})
\subset
\left(\Bmp|_{{\mathbb R}_{{\sm}_0}}\otimes{\cal S}({\mathbb R}^{3})
\right)^{\times}
\end{equation}
\end{subequations}
when the Hilbert space $\H$ in (\ref{pr}$_{\mp}$) is the space
of Lebesgue square integrable functions:
\begin{equation}
\label{hilbert}
L^{2}({\mathbb R}_{\sm_0},d\sm)\otimes L^{2}\left({\mathbb R}^3,
\frac{d^{3}\p}{2\p^0}\right)\,.
\end{equation}
The momentum operators $P^\mu$ are $\tau_{\Phi_{\pm}}$-continuous
operators given by
\begin{eqnarray}
&\!\!\!\!\!\!\!\!\!\!\!\!
\<H\phi^+|\ap\>=\gamma\sqrt{\sm}\<\phi^+|\ap\>\,,&
\<H\psi^-|\am\>=\gamma\sqrt{\sm}\<\psi^-|\am\>\nonumber\\
&& \label{values}\\
&\!\!\!\!\!\!\!\!\!\!\!\!
\<\bs{P}\phi^+|\ap\>=\sqrt{\sm}\bs{\p}\<\phi^+|\ap\>\,,&
\<\bs{P}\psi^-|\am\>=\sqrt{\sm}\bs{\p}\<\psi^-|\am\>\nonumber\\
&\!\!\!\!\!\!\!\!\!\!\!\!
\text{for all }\phi^+\in\Phi_{-} &\qquad\qquad \text{for all }\psi^-\in\Phi_{+}\,.
\nonumber
\end{eqnarray}
This follows from the special property of
$\Bmp$ 
where $\tilde{\cal S}$ has been defined such
that multiplication by $\sm^{n/2}$:
\begin{eqnarray}
\nonumber
\sm^{n/2}:\Bmp\rightarrow\Bmp\\
\label{Property2.2}
f(\sm)\rightarrow\sm^{n/2}f(\sm)
\end{eqnarray}
is $\tau_{\Phi_{\mp}}$-continuous 
for any (positive or negative) integer $n$.
The branch in \eqref{2.15.5} and \eqref{values} is chosen to be
\begin{equation}
\label{branch}
-\pi\leq {\rm Arg}\,\sm<\pi\,.
\end{equation}
Specifying the branch, even though irrelevant for the physical
values of $\sm$, is necessary  for 
obtaining the transformation
properties of $\Phi_\pm$ and
the Gamow vectors, as will be discussed in detail in 
Sections~\ref{semigroup} and~\ref{evolution}. 

It follows from the $\tau_{\Phi_\pm}$-continuity of $P^\mu$
that the conjugate operators $P^{\mu\,\times}$ in~\eqref{2.15.5} defined
by
\begin{equation}
\label{conjugate}
\begin{split}
\<\phi^+|P^{\mu\,\times}|\ap\>\equiv\<P^\mu\phi^+|\ap\>\\
\<\psi^-|P^{\mu\,\times}|\am\>\equiv\<P^\mu\psi^-|\am\>
\end{split}
\end{equation}
are everywhere defined $\tau_{\Phi_\pm^\times}$-continuous 
operators on $\Phi_{\pm}^{\times}$ ($\tau_{\Phi_{\pm}^{\times}}$
refers to the weak$^{*}$-topology of $\Phi_\pm^\times$ \cite{gelfand}). 

With the postulate \eqref{pr} \eqref{t} the wavefunctions $\<^{+}\bs{\p}j_3[\sm j]|\phi^+\>$
and $\<^{-}\bs{\p}j_3[\sm j]|\psi^-\>$
have a unique extension to the negative values of
$\sm$, $-\infty<\sm\leq (m_1+m_2)^{2}$ \cite{gadella,winter}, and
the D-L-Sch 
kets $|\amr\>$ can be analytically continued 
into the whole lower-half complex plane, cf. Section~$5$ of \cite{paper1}.
The Gamow vectors are then obtained under the requirement that 
the analytically continued $S$-matrix is  
polynomially bounded for large values of $|\sm|$. 

The derivation of the Gamow vectors $|\amr\>$ 
from the resonance poles
of the analytically continued $S$-matrix at $\sm_R=(M_R-i\Gamma_R/2)^{2}$, 
yields the following
properties of $|\amd\>$~\cite{paper1}:
\begin{enumerate}
\item The Gamow vectors $|\amd\>$ have a relativistic 
Breit-Wigner energy distribution 
and are given by the integral representation:
\begin{equation}
\label{rgv}
|\amd\>
=\frac{i}{2\pi}\int_{-\infty_{II}}^{\infty}
d\sm\frac{|\amr\>}{\sm-\sm_{R}}\,,\quad \sm_R=(M_R-i\Gamma_R/2)^{2}\,,
\end{equation}
in terms of the Dirac-Lippmann-Schwinger kets. 
Here $-\infty_ {II}$ signifies that the ``unphysical'' values
of $\sm$, $-\infty<\sm\leq 4m^2$, are in the second sheet.
The equation \eqref{rgv}  
is a relation between continuous functionals over $\Phi_+$, i.e.,
$|\amd\>\in\Phi_+^\times$.
\item The Gamow vectors are generalized eigenvectors of the mass
operator $M=(P_\mu P^\mu)^{1/2}$ and momentum operators $P^{\mu}$
with complex eigenvalues:
\begin{eqnarray}
\nonumber
\bs{P}^{\times}|\amd\>=\sqrt{\sm_R}\bs{\p}|\amd\>\\
\label{htimes}
H^\times|\amd\>=\gamma\sqrt{\sm_R}|\amd\>\\
\nonumber
M^{\times}|\amd\>=\sqrt{\sm_R}|\amd\>\,.
\end{eqnarray}
\item
The Gamow vectors are elements of a complex basis system for the in-states.
This means that the prepared in-state vector
$\phi^+\in\Phi_-$ can be represented as
\begin{equation}
\label{complexbasisexpansion}
\phi^+ = \phi^{bg}+\sum_{i=1}^{N}|\sm_{R_i}^-\>c_{R_i}\,,\quad \text{where }
c_{R_i} = (2\pi R^{(i)}/i)\<^+\sm_{R_i}|\phi^+\>\,,
\end{equation}
where $N$ is the number of resonance poles in the $j$-th partial wave amplitude
($N=2$ in case of \cite{paper1} Figure $2$).
In this way the in-state $\phi^+$ has been decomposed into a vector representing 
the non-resonant part $\phi^{bg}$ $(5.20)$ of \cite{paper1} and a sum over the
$N$ Gamow vectors each representing a resonance state. The complex eigenvalue
expansion \eqref{complexbasisexpansion} is an alternative generalized eigenvector expansion
to the Dirac's eigenvector expansion \eqref{t1}.

While \eqref{t1} expresses the in-state $\phi^+$ in terms of the
Lippmann-Schwinger kets $|\sm^+\>\in\Phi_-^\times$, which are generalized eigenvectors
of the mass operator $P_\mu P^\mu$ with real eigenvalue $\sm$, \eqref{complexbasisexpansion}
is an expansion of $\phi^+\in\Phi_+^\times$ in terms of eigenkets $|\sm_{R_i}^-\>\in\Phi_+^\times$
of the same {\em self-adjoint} mass operator $P_\mu P^\mu$ with complex generalized eigenvalue
$\sm_{R_i}=(M_{R_i}-i\Gamma_{R_i}/2)^2$ and the vector $\phi^{bg}$. The term $\phi^{bg}$ is
defined by $(5.20)$ of \cite{paper1} and is therefore an element of $\Phi_+^\times$.
We can rewrite $(5.20)$ of \cite{paper1} into a familiar form. According to the van Winter
theorem \cite{winter}, a Hardy class function on the negative real axis is uniquely
determined by its values on the real positive axis (cf. Appendix~B of \cite{paper1}).
Therefore one can use the Mellin transform to rewrite the integral on the l.h.s. of 
$(5.20)$ of \cite{paper1} into an integral over the interval $m_0^2\leq\sm<\infty$ and obtain
\begin{eqnarray}
\nonumber
\<\psi^-|\phi^{bg}\> &=& \int_{\sm_0}^{-\infty_{II}}d\sm
\<\psi^-|\sm^-\>S_j(\sm)\<^+\sm|\phi^+\>\\
\label{n1.21}
&=&\int_{\sm_0}^{\infty}d\sm\<\psi^-|\sm^-\>b_j(\sm)\<^+\sm|\phi^+\>\,,
\end{eqnarray}
where $b_j(\sm)$ is uniquely defined by the values of $S_j(\sm)$ on the negative
real axis. Without more specific information about $S_j(\sm)$, we cannot be certain
about the energy dependence of the background $b_j(\sm)$. If there are no further
poles or singularities besides those included in the sum, then $b_j(\sm)$ is likely
to be a slowly varying function of $\sm$ \cite{gadella2}. Using \eqref{n1.21}, omitting the arbitrary
$\psi^-\in\Phi_+$, we write the complex basis vector expansion \eqref{complexbasisexpansion}
of the prepared in-state vector $\phi^+$ as:
\begin{equation}
\label{n1.22}
\phi^+ = \sum_{i=1}^{N}|\sm_{R_i}^-\>c_{R_i}+\int_{\sm_0}^{\infty}d\sm\,|\sm^-\>
\<^+\sm|\phi^+\>b_j(\sm)\,;\quad |\sm_{R_i}^-\>\,,\,|\sm^-\>\in\Phi_+^\times
\end{equation}
which is a functional equation over the space $\Phi_+$.

The basis vector expansion \eqref{n1.22} shows that the resonances
appear here on the same footing as the bound states in the usual basis vector expansion
for a system with discrete energy eigenvalue, with the only difference that the
bound states are represented by proper vectors $|E_n)\in\H$ and the Gamow states are
represented by generalized vectors $|\sm_{R_i}^-\>\in\Phi_+^\times$. The basis
vector expansion \eqref{n1.22} shows that in addition to the superposition of $N$
Gamow states, there appears an integral (or continuous superposition) over the continuous
basis vectors $|\sm^-\>$ with a weight function $b(\sm)\<^+\sm|\phi^+\>$, where
the wave function $\phi^+(\sm)=\<^+\sm|\phi^+\>$ depends upon the particular preparation
of the state $\phi^+$ and will change with the preparation from experiment to experiment.

Since the complex basis vector expansion \eqref{n1.22} is such an important formula,
we want to give it here also in the un-abbreviated notation $|\sm^-\>\rightarrow
|\bs{\p}j_3[\sm j]^-\>$ corresponding to the form \eqref{t1} for the
Dirac basis vector expansion. The vector $\phi^+$ has a velocity distribution described
by the well-behaved (Schwartz) function of $\bs{\p}$:
$$
f_j(\bs{\p}) = f(\bs{\p}) = \<^+\bs{\p}j_3[\sm j]|\phi^+\> \in {\cal S}({\mathbb R}^{3})
$$
and to each resonance pole corresponds the space of Gamow vectors $(5.29)$ of \cite{paper1}
(one for every $f(\bs{\p}) \in {\cal S}({\mathbb R}^3)$:
\begin{equation}
\label{4.10.8}
\phi_{j \sm_{R_{i}}}^G=|[\sm_{R_i} j]f^-\>=\sum_{j_3}\int\frac{d^3\p}{2\p^0}
|\bs{\p}j_{3}[\sm_{R_{i}}j]^-\>c_{R_i}f_{j_3}(\bs{\p})
\end{equation}
Each of these Gamow vectors
represents a $4$-velocity or momentum 
wave-packet of the unstable particle characterized
by $\sm_{R_{i}}$, $j$. 
In addition to the superposition of
Gamow vectors in \eqref{n1.22} 
the in-state $\phi^+$ 
also contains a non-resonant background vector  
$|B\>$ which describes the  
non-resonant background,
\begin{equation}
\label{n1.24}
|B_f\> = \int d\sm \int \sum_{j_3} \frac{d^3\p}{2\p^0}|\bs{\p}j_3[\sm j]^-\>b_j(\sm)f(\bs{\p})\,.
\end{equation}
This vector corresponds to the second term on the r.h.s.\ of \eqref{n1.22}. The complex
basis vector expansion of every prepared in-state vector $\phi^+\in\Phi_-$ with momentum
distribution described by $f(\bs{\p})\in{\cal S}(\mathbb{R}^3)$ is thus given by
\begin{equation}
\label{n1.25}
\phi^+ = \sum_{i=1}^{N}\phi_{j\sm_{R_i}}^{G} + |B_f\> = \sum_{i=1}^{N}|[\sm_{R_i},j]f^-\> + |B_f\>\,,
\end{equation}
where the terms in the sum are defined by \eqref{4.10.8} and \eqref{n1.24}.

\hspace*{0.5cm}The complex basis vector expansion \eqref{n1.22}, \eqref{n1.25} is an exact 
consequence of the new hypothesis \eqref{pr} \eqref{t}.
Thus, representing
the in-state $\phi^+$ by a superposition of Gamow vectors by omitting
$|B\>$ in \eqref{n1.22} \eqref{n1.25}, as is often done in the ``effective theories''
of resonances and decay, is an approximation. It corresponds to the Weisskopf-Wigner
approximation \cite{weisskopf}.
\end{enumerate}

\section{Action of $U(\l,x)$ on $\Phi_\pm$}\label{semigroup}
In \cite{paper1} the Gamow kets $|\bs{\p}j_3[\sm_Rj_R]^-\>$ $(5.28)$
of \cite{paper1}, and the
resonance state vectors  ($(5.29)$ of \cite{paper1}) 
were defined from the pole of
the $S$-matrix.  To obtain the relativistic Gamow
vectors we needed,  in addition to the standard analyticity assumption of the
$S$-matrix, the analyticity and smoothness assumption \eqref{t} of the
energy wave functions, i.e., the new hypothesis \eqref{pr}. In Section
\ref{evolution} we shall derive the transformation property of the
Gamow vectors under Poincar\'e transformations $U(\Lambda,x)$. 
As a preparation for this derivation, we consider in this section the effect of the hypothesis
\eqref{pr}, \eqref{t} on the transformation 
properties the D-L-Sch kets $|\bs{\p}j_3[\sm
j]^\mp\>\in\Phi_\pm^\times$. We shall see that, if the  D-L-Sch kets
are mathematically well defined as functionals on the Hardy spaces
$\Phi_\pm$, then their transformations will not be defined for the
whole Poincar\'e group but only for the two semigroups into the
forward and backward light cones. This is in contrast to what is
usually assumed for the (mathematically not defined) plane wave
solutions of the Lippmann-Schwinger equations \cite{weinberg}.
Since $(\Lambda,x)=(I,x)(\Lambda,0)$,
we start by considering the action of space-time translations by a 
$4$-vector $x$, $U(I,x)=e^{iP.x}$ on the space of observables
$\Phi_+$, and of $U^\dagger (I,x) = e^{-iP.x}$ restricted to the space
of states $\Phi_-$.

The space-time translation by a $4$-vector $x$ 
of any out-observable $\psi^-\in \Phi_+$
defined by a registration apparatus (detector)
is given by~: $U(I,x)\psi^-=e^{iP.x}\psi^-$.
The vector $U(I,x)\psi^-$ is realized (in the sense of \eqref{t3})
by the wavefunction
\begin{equation}
\label{1}
\begin{split}
\<U(I,x)\psi^{-}|\am\>&=\<\psi^-|U^{\times}(I,x)|\am\>\\
&=\<\psi^{-}|e^{-ix^{\mu}P_{\mu}^{\times}}|\am\>\\
&=\<\psi^{-}|e^{-i\left[H^{\times}t-\bs{x}.\bs{P}^{\times}\right]}|\am\> \\
&=e^{-i\gamma\sqrt{\sm}(t-\bs{x}.\bs{v})}\<\psi^-|\am\>\\
&=e^{-ipx}\<\psi^-|\hat{\bs{p}}j_3[\sm j]^-\>
\end{split}
\end{equation}
where \eqref{2.15.5} has been used. 
Equation~\eqref{1} is regarded as the defining 
formula for $U(I,x)\psi^-$. 
It is inferred here by formally using the conventional Dirac bra-ket
formalism, with the difference that we write the conjugate operator
$U^\times$, 
which is the extension of
$U^{\dagger}$ to $\Phi_+^\times$ 
rather than writing the Hilbert space adjoint operator
$U^\dagger=U^\times|_{\H}$ as is common in the standard 
literature, e.g, \cite{joos,weinberg,wigner}.
This distinction between $U^\dagger$ and $U^\times$ is particular to the
Rigged Hilbert Space formulation of the Dirac ket formalism,
and the extension $U^\times$ depends upon the choice of the spaces $\Phi$.
It is a different operator for the space $\Phi$ of the Rigged Hilbert
Space of footnote~\ref{f1} than for $\Phi_+$ or for $\Phi_-$.
We will obtain now the conditions under which $U(I,x)$, 
defined by~\eqref{1},
is a $\tau_{\Phi_+}$-continuous operator on $\Phi_+$,  
i.e., 
$\psi^{-}\mapsto U(I,x)\psi^{-}$ is a continuous map from 
$\Phi_+\rightarrow\Phi_+$, and $U^\times(I,x)$ can be defined as a continuous
operator in $\Phi_+^\times$, [\ref{paper1}, Appendix~A].

To establish the continuity of $U(I,x)$ on $\Phi_+$, we consider
first the invariance of $\Phi_+$ under $U(I,x)$.
Since, as seen in \eqref{1}, the action of $U(I,x)$ on $\Phi_+$ is a
multiplication by $e^{-i\gamma\sqrt{\sm}(t-\bs{x}.\bs{v})}$, we have to 
find the conditions 
under which the statement
\begin{align}
\nonumber
\<\psi^-|\am\>\in\left.\Bm\right|_{\mathbb{R}_{\sm_{0}}}
\otimes {\cal S}({\mathbb{R}^{3}}) \Longrightarrow \\
e^{-i\gamma\sqrt{\sm}(t-\bs{x}.\bs{v})}\<\psi^-|\am\>
\in\left.\Bm\right|_{\mathbb{R}_{\sm_{0}}}
\otimes {\cal S}({\mathbb{R}^{3}})\label{p4}
\end{align}
is true.
Since $$e^{-i\gamma\sqrt{\sm}(t-\bs{x}.\bs{v})}
=e^{-i\sqrt{\sm}(\sqrt{1+\bs{\p}^{2}}t-\bs{x}.\bs{\p})}\,,$$
the Schwartz property in the $\bs{\p}$ variable is satisfied. Also, 
despite the appearance of $\sqrt{\sm}$ in~\eqref{1}
and~\eqref{p4} the smoothness requirement in the variable $\sm$ 
is also preserved
by virtue of \eqref{Property2.2}.
Thus, so far
$$
\<U(I,x)\psi^-|\am\>\in\tilde{\cal S}|_{{\mathbb R}_{\sm_{0}}}\otimes
{\cal S}({{\mathbb R}^{3}})\,.
$$
We shall now investigate the analyticity property 
of $\<U(I,x)\psi^-|\am\>$ to determine whether and for which $(I,x)$ they
are Hardy
class functions from below for the variable $\sm$ (Definition~\ref{h:1}).

To apply Definition~\ref{h:1} 
of ${\cal H}_{-}^{2}$ to $\<U(I,x)\psi^{-}|\am\>$,
we consider the behavior of $\sqrt{\sm}$ for the chosen 
branch~\eqref{branch}
$-\pi\leq{\rm Arg}\,\sm<\pi$. 

Let $\sm=\sigma+i\eta$.
If $\eta<0$ then $-\pi< {\rm Arg}\,{\sm}<0$, hence 
$-\frac{\pi}{2}< \frac{{\rm Arg}\,\sm}{2}<0$.
Therefore~:
\begin{equation}
\label{s-}
\begin{split}
\sqrt{\sm}&=(\sigma^{2}+\eta^{2})^{1/4}
\left[\cos\frac{{\rm Arg}\,\sm}{2}+i\sin\frac{{\rm Arg}\,\sm}{2}\right]\\
&=(\sigma^{2}+\eta^{2})^{1/4}\left[\left|\cos\frac{{\rm Arg}\,\sm}{2}\right|
-i\left|\sin\frac{{\rm Arg}\,\sm}{2}\right|\right]\, .
\end{split}
\end{equation}
We see from \eqref{s-} that if $\sqrt{\sm}=a+ib$, then $a\geq 0$, $b\leq 0$, 
so that:
\begin{equation}
\label{s--} 
\sqrt{\sm}=|a|-i|b|\quad\text{for }{\rm Im}\,\sm=\eta<0\, .
\end{equation}
Similarly one can see  
\begin{equation}
\label{s++}
\sqrt{\sm}=|a|+i|b|\quad\text{for }{\rm Im}\,\sm=\eta>0\, .
\end{equation}

With~\eqref{s--}, we 
test $\<U(I,x)\psi^-|\am\>$ given by \eqref{1}
against the defining criterion \eqref{h1} for $\H_{-}^{2}$:
\begin{eqnarray}
\lefteqn{\underset{\eta<0}{\rm sup}\int_{-\infty}^{\infty}
\left|e^{-i\gamma(t-\bs{x}.\bs{v})\sqrt{\sm}}\<\psi^{-}|\am\>\right|^{2}
d\sigma}\nonumber\\
& &=\underset{\eta<0}{\rm sup}\int_{-\infty}^{\infty}
\left|e^{-i\gamma(t-\bs{x}.\bs{v})(|a|-i|b|)}\<\psi^{-}|\am\>\right|^{2}
d\sigma\nonumber\\
& &=\underset{\eta<0}{\rm sup}\int_{-\infty}^{\infty}
e^{-2\gamma(t-\bs{x}.\bs{v})|b|}
\left|\<\psi^{-}|\am\>\right|^{2}d\sigma \nonumber\\
& &=\underset{\eta<0}{\rm sup}\int_{-\infty}^{\infty}
e^{-2\gamma(t-\bs{x}.\bs{v})
(\sigma^{2}+\eta^{2})^{1/4}
|\sin\frac{{\rm Arg}\,(\sigma+i\eta)}{2}|}\left|\<\psi^{-}|\am\>
\right|^{2}d\sigma\,.\label{3}
\end{eqnarray}
The exponential in~\eqref{3} is bounded for all $\eta<0$ if
\begin{equation}
\label{4}
t-\bs{x}.\bs{v}\geq 0\, .
\end{equation}
Moreover, there exist ${\cal H}_{-}^{2}$ functions $\<\psi^-|\am\>$
such that~\eqref{3} is not bounded for 
$t-\bs{x}.\bs{v}\leq 0$ (Proposition~\ref{theb}). 
Hence,
\begin{subequations}
\begin{eqnarray}
\nonumber
&\!\!\!\!\!\!\<U(I,x)\psi^{-}|\am\>\in \bm 
\text{ for all }\<\psi^{-}|\am\>\in \bm\\
&\text{if and only if }\,\,\, t-\bs{x}.\bs{v}\geq 0\,.\label{5}
\end{eqnarray}
Similarly, using~\eqref{s++}, it can be shown that:
\begin{eqnarray}
\nonumber
&\!\!\!\!\!\!\<U(I,x)\phi^{+}|\ap\>\in \bp 
\text{ for all }\<\phi^{+}|\ap\>\in \bp\\
&\text{if and only if }\,\,\, 
t-\bs{x}.\bs{v}\leq 0\, .\label{6}
\end{eqnarray}
\end{subequations}
The relation \eqref{5} 
indicates that a necessary condition 
for $\Phi_+$ 
to be invariant under $U(I,x)$, i.e,  
$\psi^{-}\rightarrow U(I,x)\psi^{-}\in\Phi_{+}$,
is that for any given $4$-vector $x$,
\begin{subequations}
\label{eitheror}
\begin{equation}
\label{either}
t-\bs{x}.\bs{v}\geq 0\,\,\,\text{ for all }\,\,\,|\bs{v}|\leq 1\,.
\end{equation}
Similarly, \eqref{6} 
indicates that $U(I,x)$ leaves $\Phi_-$ invariant if and only if
\begin{equation}
\label{or}
t-\bs{x}.\bs{v}\leq 0\,\,\,\text{ for all }\,\,\,|\bs{v}|\leq 1\,.
\end{equation}
\end{subequations}

The conditions of \eqref{eitheror} ensure
the continuity of $U(I,x)$ on $\Phi_\pm$ (Appendix~\ref{o}).
The four vectors $(t,\bs{x})$ which fulfill either~\eqref{either}
or~\eqref{or} have the property $x^2\geq0$, and thus we shall refer to
transformation vectors that fulfill \eqref{eitheror} 
as causal space-time
translations. Therefore, for
causal space-time translations,
the conjugate operator of $U(I,x)|_{\Phi_+}$, 
$\left( U(I,x)|_{\Phi_+}\right)^{\times}$,
is only defined for $x^2\geq0$ and $t\geq 0$, and 
the conjugate operator of $U(I,x)|_{\Phi_-}$,
$\left( U(I,x)|_{\Phi_-}\right)^{\times}$, is only defined for
$x^2\geq0$ and  $t\leq 0$.
Note that $U^\times_+(I,x)=\left(U(I,x)|_{\Phi_{+}}\right)^{\times}$ and 
$U^\times_-(I,x)=\left(U(I,x)|_{\Phi_{-}}\right)^{\times}$ are uniquely defined
extensions of the Hilbert space adjoint operator $\left(U(I,x)\right)^\dagger$
to the spaces $\Phi_+^\times$ and $\Phi_-^\times$ respectively, cf.,
eq. $(A5)$ of~\cite{paper1}.
Thus, for any causal $(I,x)$
\begin{subequations}
\label{p5}
\begin{equation}
\label{p5.1}
\tag{\ref{p5}$_{+}$}
U_+^{\times}(I,x)|\am\>=e^{-i\gamma \sqrt{\sm}(t-\bs{x}.\bs{v})}|\am\>\,,
\text{ only for }t\geq 0,\ x^2\geq0
\end{equation}
\begin{equation}
\label{p5.2}
\tag{\ref{p5}$_{-}$}
U_-^{\times}(I,x)|\ap\>=e^{-i\gamma \sqrt{\sm}(t-\bs{x}.\bs{v})}|\ap\>\,,
\text{ only for }t\leq 0,\ x^2\geq0
\end{equation}
\end{subequations}

According to~\eqref{eitheror}, the spaces of in-states $\Phi_-$ and of 
out-observables $\Phi_+$  remain invariant under causal space-time
translations with $t\leq 0$ and $t\geq 0$ respectively. Furthermore,
since proper orthochronous Lorentz transformations preserve the
property $x^2\geq0$ as well as the sign of $t$, we see that the set
\begin{subequations}
\label{2.11}
\begin{equation}
\label{2.11+} 
\tag{\ref{2.11}$_{+}$}
{\P}_+\equiv\{(\Lambda,x):\ \det\Lambda=1,\ \Lambda^0_{\ 0}\geq1,\
x^2\geq0,\ t\geq0\}\nonumber
\end{equation}
leaves the space $\Phi_+$ invariant under $U_+(\Lambda, x)$. This is  
the causal Poincar\'e semigroup
into the forward light cone. Similarly, the causal Poincar\'e
semigroup into the backward light cone can be defined as
\begin{equation}
\label{2.11-} 
\tag{\ref{2.11}$_{-}$}
{\P}_-=\{(\Lambda,x):\ \det\Lambda=1,\ \Lambda^0_{\ 0}\geq1,\ 
x^2\geq0,\ t\leq0\}\nonumber
\end{equation}
\end{subequations}
It leaves the space $\Phi_-$ invariant under $U_-(\Lambda,x)$. 
What~\eqref{p5} infers (Appendix~\ref{o}) is that the map 
\begin{subequations}
\label{2.12}
\begin{equation}
\label{2.12+} 
\tag{\ref{2.12}$_{+}$}
U_+(I,x)\Phi_+\rightarrow\Phi_+
\end{equation}
\parbox[t]{\textwidth} 
{is $\tau_{\Phi_+}$-continuous only when $x^2\geq0,\ t\geq0$, i.e.,
when $(I,x)\in{\P}_+$, and that the map}
\begin{equation}
\label{2.12-} 
\tag{\ref{2.12}$_{-}$}
U_-(I,x)\Phi_-\rightarrow\Phi_-
\end{equation} 
\end{subequations}
is $\tau_{\Phi_-}$-continuous only when $x^2\geq0,\ t\leq0$, i.e.,
when $(I,x)\in{\P}_-$. 

For $t\lessgtr0$ the  operators
$U_\pm(1,x)$ are not $\tau_{\Phi_\pm}$-continuous and the
conjugate operators $U_\pm^\times(1,x)$ are not defined for
$t\lessgtr0$.

That is, the space-time translation subgroup $\{(I,x)\}\subset{\cal
P}$, where $x$ is any four vector, time-like or space-like and
$-\infty<t<\infty$, which is represented by a group of unitary and
thus $\tau_\H$-continuous operators $\{U(I,x)\}$ in the Hilbert space
$\H$, has two subsemigroups 
\begin{subequations}
\label{2.13}
\begin{equation}
\label{2.13+} 
\tag{\ref{2.13}$_{+}$}
\{(I,x):\ x^2\geq0,\ t\geq0\}\subset{\cal{P}}_+
\end{equation}
\text{and}
\begin{equation}
\label{2.13-} 
\tag{\ref{2.13}$_{-}$}
\{(I,x):\ x^2\geq0,\ t\leq0\}\subset{\cal{P}}_-
\end{equation}
\end{subequations}
represented by the $\tau_{\Phi_\pm}$-continuous operators
$U_\pm(I,x)=U(I,x)|_{\Phi_\pm}$.
It should be noted that the topology
$\tau_{\Phi_\pm}$ under which $U_\pm(I,x)$ are continuous operators is
the topology of the Hardy spaces (\ref{t}$_\pm$) with respect to which
the generators of the Poincar\'e  
transformations are already $\tau_{\Phi_\pm}$-continuous operators, as needed
for the Dirac bra and ket formalism.  

Having obtained \eqref{p5} 
for the action of space-time translations, we now 
consider the action of Lorentz transformations $U(\Lambda,0)$
on the spaces $\Phi_{\pm}$. From the conventional Dirac bra-ket
formalism \cite{wightman,joos,macfarlane,weinberg,wigner}, or when the
$\<\psi^-|\bs{\p}j_3[\sm j]^-\>$ are considered as Lebesgue square
integrable functions in the Hilbert space \eqref{hilbert}, 
$U(\Lambda,0)\psi^-$ is given by:
\begin{align}
\nonumber
\<U(\Lambda,0)\psi^-|\am\>&=\sum_{j'_{3}}D^{j}_{j'_{3}j_{3}}(W(\Lambda^{-1},p))
\<\psi^-|\bs{\Lambda}^{-1}\bs{\p}j'_{3}[\sm j]^-\>\\
&=\<\psi^-|U^{\times}(\Lambda,0)|\am\>\,.
\label{p6}
\end{align}
where $W(\l,p)=L^{-1}(\l p)\l L(p)$ is the Wigner rotation, and
$D^{j}$ is the rotation matrix corresponding to the $j$-th
angular momentum.
This is taken as the definition of $U(\l,0)\psi^{-}$ 
and of the conjugate operator 
$U^\times(\Lambda)\supset U^\times(\Lambda)|_{\H}=U^\dagger(\Lambda)$
that satisfies the multiplication law: 
$U(\l_1,0)U(\l_2,0)=U(\l_1\l_2,0)$. In \eqref{p10.1.1} below,
it will be shown that this definition agrees with the heuristic transformation
properties of the Dirac kets of a Wigner representation~\eqref{v25}.
Since the rotation matrix $D^j$
in~\eqref{p6} is a polynomial in its parameters, \eqref{p6}
defines $U(\Lambda,0)$ as a $\tau_{\Phi_+}$-continuous operator
on $\Phi_+$. Hence, $U^{\times}(\Lambda,0)$
in~\eqref{p6} is a well-defined operator on $\Phi_{+}^{\times}$.
Thus, omitting the arbitrary $\psi^-$, we write \eqref{p6}  as an
equation between the functionals  
$|\bs{\p}j_3[\sm j]^-\>\in\Phi_+^\times$
\begin{equation}
\label{p8}
U^{\times}(\Lambda,0)|\am\>=
\sum_{j'_{3}}D^{j}_{j'_{3}j_{3}}(W(\Lambda^{-1},p))
|\bs{\Lambda}^{-1}\bs{\p}j_3' [\sm j]^{-}\>
\end{equation}
This agrees with the standard formula for
$U^\dagger(\Lambda,0)=U(\Lambda^{-1},0)$ of the Wigner
representations. The homogeneous Lorentz transformations $(\Lambda,0)$
are also here unitarily represented and the $U^\times(\Lambda,0)$ form
a group. 
Combining~\eqref{1} and~\eqref{p6}, we obtain:
\begin{subequations}
\label{p9}
\begin{equation}
\!\!\!\!\!\!
\<U(\Lambda,x)\psi^-|\am\>=e^{-i p.x}
\sum_{j'_{3}}D^{j}_{j'_{3}j_{3}}(W(\Lambda^{-1},p))
\<\psi^-|\bs{\Lambda}^{-1}\bs{\p}j'_{3}[\sm j]^{-}\>,\ 
t\geq0,\ x^2\geq0
\label{p9minus}\tag{\ref{p9}$_{-}$}
\end{equation}
Expressions analogous to \eqref{p6} and \eqref{p8}, which are
obtained for $\psi^-\in\Phi_+$ and $|\am\>$, apply also for
$\phi^+\in\Phi_-$ and $|\ap\>$.
Hence,
\begin{equation}
\label{p9plus}\tag{\ref{p9}$_{+}$}
\!\!\!\!\!\!
\<U(\Lambda,x)\phi^+|\ap\>=e^{-i p.x}
\sum_{j'_{3}}D^{j}_{j'_{3}j_{3}}(W(\Lambda^{-1},p))
\<\phi^+|\bs{\Lambda}^{-1}\bs{\p}j'_{3}[\sm j]^{+}\>,\   
t\leq0,\ x^2\geq0
\end{equation}
\end{subequations}
It is straightforward to check that~\eqref{p9} satisfies the 
multiplication law
$$U(\l_1,x_1)U(\l_2,x_2)=U(\l_1\l_2,\l_1 x_2+x_1)\,.$$
The transformations (\ref{p9}$_\pm$) 
we write again as functional equations in $\Phi_\mp^\times$.
Combining~(\ref{p5}$_\pm$) and~\eqref{p8}, we obtain
\begin{subequations}
\label{p10}
\begin{equation}
\label{p10-}\tag{\ref{p10}$_{-}$}
U_+^{\times}(\Lambda,x)|\am\>
=e^{-i\,p.x}\sum_{j'_{3}}D^{j}_{j'_{3}j_{3}}(W(\Lambda^{-1},p))
|\bs{\Lambda}^{-1}\bs{\p}j'_{3}[\sm j]^{-}\>,\quad t\geq 0,\ x^2\geq0
\end{equation} 
\text{And similarly we obtain for $|\bs{\p}j_3[\sm
j]^+\>\in\Phi_-^\times$}
\begin{equation}
\label{p10+}\tag{\ref{p10}$_{+}$}
U_-^{\times}(\Lambda,x)|\ap\>
=e^{-i\,p.x}\sum_{j'_{3}}D^{j}_{j'_{3}j_{3}}(W(\Lambda^{-1},p))
|\bs{\Lambda}^{-1}\bs{\p}j'_{3}[\sm j]^{+}\>,\quad t\leq 0, x^2\geq0 
\end{equation}
\end{subequations}
Here
$U^{\times}(\Lambda,x)=\left(U(I,x)U(\Lambda,0)\right)^{\times}
=U^\times(\Lambda,0)U^\times(I,x)$, 
and $(\Lambda,x)\in{\P}_{\pm}$.
Equations~\eqref{p9} and~\eqref{p10} express the transformation
properties of $\Phi_\pm$ and their basis vectors $|\amp\>$
under $(\Lambda,x)\in{\P}_{\pm}$.

In order to show that~\eqref{p10} has the same appearance as 
the standard expressions
for the action of the $U(\Lambda,x)$ on the Dirac kets of the unitary
Wigner representation, we consider the representation
$U^{\times}(\l^{-1},-\l^{-1} x)=U^{\times}((\Lambda,x)^{-1})$ of the
inverse element $(\Lambda,x)^{-1}$. According to~\eqref{p10}, we obtain
\begin{eqnarray}
&U^{\times}(\l^{-1},-\l^{-1}x)|\amp\>
=e^{i\l p.x}\sum_{j_{3}'}D^{j}_{j_{3}'j_{3}}(W(\l,p))
|\bs{\l\p}j_{3}'[\sm j]^{\mp}\>
\nonumber\\
&\text{only for $t\leq 0$ (for $-$), $t\geq 0$ (for $+$)}\,.
\label{p10.1.1}
\end{eqnarray}
Since $(U(\l,x)^{-1})^\times=U^\times(\l^{-1},-\l^{-1}x)$ is the 
extension of the Hilbert space operator 
$U^{\dagger}((\l,x)^{-1})=U^{-1}((\l,x)^{-1})=U(\l,x)$, we
would formally (i.e., if we would not distinguish between
$U^\dagger$ and $U^{\times}$, and between
$|\a\>$ and $|\amp\>$) write~\eqref{p10.1.1} as
\begin{equation}
\label{v25}
\text{``$U(\l,x)$''}|\a\>=e^{i\l p.x}\sum_{j_{3}'}D^{j}_{j_{3}'j_{3}}(W(\l,p))
|\bs{\l\p}j_{3}'[\sm j]\>\,.
\end{equation}
This is the standard formula for the transformation of the Dirac basis
kets of a Wigner representation \cite{joos,weinberg}.  It is assumed to hold for all
$(\Lambda,x)\in\P$. 

In the standard treatment of scattering theory, 
the Dirac kets $|\a\>$ (or the momentum eigenkets 
$|\bs{p}j_3[\sm j]\>$
which one almost always uses) are mathematically not fully 
defined, i.e., one does 
not define the space $\Phi\subset\H$ of which they are functionals. 
If one chooses for $\Phi$ the Schwartz space defined in footnote$^1$, 
i.e., its nuclear topology is given by the countable norms
$(\psi,(\Delta+1)^{p}\phi)$, $p=0,1,2,\cdots$ where $\Delta$ is 
the Nelson operator, and if one defines the momentum kets as
functionals, $|\bs{\p}j_3[\sm j]\>\in\Phi^\times$, then
``$U(\Lambda,x)$'' in \eqref{v25} can be defined as
$\left(U(\Lambda,x)^{-1}|_\Phi\right)^\times$, the conjugate operator
in $\Phi^\times$. In this space $\Phi^\times$
\eqref{v25} is indeed a representation of the whole group $\P$.  
($\Phi$
is  the space of differentiable vectors of the unitary representation
$[m,j]$).  In this Rigged Hilbert Space 
$\Phi\subset\H[m,j]\subset\Phi^{\times}$, one does not have semigroup 
representations and time asymmetry, the space
$\Phi$ (of differentiable vectors) is invariant with respect
to the transformations $(\Lambda,x)$. 
But the space $\Phi^\times$ does not contain the plane wave solutions of
the Lippmann-Schwinger equation because of the (infinitesimal)
imaginary part $\mp i\epsilon$ of the energy (and therefore of
$\sm$). 

For time asymmetry given by the 
semigroup $\P_\mp$ one requires
the Hardy Rigged Hilbert Spaces (\ref{pr}$_{\mp}$), 
for which the $\Phi_\mp^\times$ are larger than the space
$\Phi^\times$. These spaces $\Phi_{\mp}^{\times}$ 
contain
the plane wave solutions of the Lippmann-Schwinger equation
$|\bs{p}j_{3}[\sm j]^{\pm}\>$. In addition, they 
also contain the continuation of the Lippmann-Schwinger kets 
to the whole lower or upper complex half-plane. In particular, they contain
the relativistic Gamow vectors which are defined 
by integrals of the Lippmann-Schwinger kets with Cauchy kernels around the
resonance poles 
of the $S$-matrix~\cite{paper1}. 
\begin{equation}
\label{rgv'} 
\tag{\ref{rgv}'}
|\bs{\p}j_3[\sm_Rj_R]^-\>=-\frac{i}{2\pi}\oint 
 d\sm\frac{|\bs{\p}j_3[\sm j_R]^-\>}{\sm-\sm_R}
\end{equation}
Under the Hardy space assumption \eqref{pr}, i.e., considered as
elements of the space $\Phi_+^\times$, these relativistic Gamow vectors
acquire the representation \eqref{rgv} with a Breit-Wigner energy
distribution. In the following section we make use of the results
obtained in the present section to
derive the action of $U(\Lambda,x)$ on the relativistic Gamow vectors 
$|\amd\>$.

\section{Transformation Properties of the 
Relativistic Gamow Vectors}\label{evolution}

We first obtain the transformation properties for 
space-time translations of the relativistic Gamow
vector $|\amd\>$ using their integral representation~\eqref{rgv} in
terms of the Lippmann-Schwinger kets. Taking the functional
of~\eqref{rgv} at an  
arbitrary vector $U(I,x)\psi^-$, we obtain
\begin{align}
\nonumber
\<U(I,x)\psi^{-}|\amd\>&=\frac{i}{2\pi}\int_{-\infty}^{\infty}
d\sm\frac{\<U(I,x)\psi^{-}|\amr\>}{\sm-\sm_{R}}\nonumber\\
&=\frac{i}{2\pi}\int_{-\infty}^{\infty}
d\sm\frac{e^{-i\gamma\sqrt{\sm}(t-\bs{x}.\bs{v})}\<\psi^{-}|\amr\>}
{\sm-\sm_{R}}\,, \label{11}
\end{align}
where~\eqref{1} is used to obtain the second equality.
According to~\eqref{5}, the numerator of the integrand
in~\eqref{11} 
is in ${\cal H}_{-}^{2}$ for all $\psi^-\in\Phi_+$ if and only
if \eqref{either} is fulfilled, i.e., if and only if  
$t\geq 0,\ x^2\geq0$. Hence, we can apply the Titchmarsh theorem 
(Cf.~B.1, Appendix B of \cite{paper1}) to the function
$e^{-i\gamma\sqrt{\sm}(t-\bs{x}\bs{v})}\<\psi^-|\amr\>$ and obtain 
\begin{eqnarray}
\label{12}
&\<U(I,x)\psi^{-}|\amd\>=e^{-i\gamma\sqrt{\sm_{R}}(t-\bs{x}.\bs{v})}
\<\psi^{-}|\amd\>\\
&\nonumber \text{ if and only if}\ x^2\geq0,\ t\geq 0
\end{eqnarray}
Equation~\eqref{12}, being valid for all $\psi^-\in\Phi_+$, is written
as the
generalized eigenvalue equation for $|\amr\>\in\Phi_+^\times$ 
\begin{eqnarray}
\!\!\!\!\!\!
U(I,x)^{\times}|\amd\>&=&e^{-ix.P^{\times}}|\amd\>\nonumber\\
\label{13}
&=&e^{-i\gamma\sqrt{\sm_{R}}(t-\bs{x}.\bs{v})}|\amd\>
\text{ if and only if}\  x^2\geq0, t\geq 0\nonumber\\
\end{eqnarray}
Equation~\eqref{13} shows that the Gamow vector $|\amd\>$
is a generalized eigenvector for $U(I,x)$ {\it only } for 
space-time translations into the forward light cone.

In the same way, to obtain the action of $U(\Lambda,0)$ on
$|\amd\>$, we apply~\eqref{rgv} to $U(\Lambda,0)\psi^-$~:
\begin{align}
\nonumber
\!\!\!\!\!\!
\<U(\Lambda,0)\psi^-|\amd\>&=\frac{i}{2\pi}\int_{-\infty}^{\infty}d\sm
\frac{\<U(\l,0)\psi^-|\amr\>}{\sm-\sm_{R}}\nonumber\\
&=\frac{i}{2\pi}\sum_{j_3'}D^{j_R}_{j_3'j_3}(W(\l^{-1},p))
\int_{-\infty}^{\infty}d\sm
\frac{\<\psi^-|\bs{\l}^{-1}\bs{\p}j_3'[\sm j_R]^{-}\>}
{\sm-\sm_{R}}\nonumber\\
&=\sum_{j_3'}D^{j_R}_{j_3' j_3}(W(\l^{-1},p))\<\psi^-|
\bs{\l}^{-1}\bs{\p} j_3'[\sm_{R} j_{R}]^{-}\>\,.
\label{13.1}
\end{align}
We used in~\eqref{13.1} the crucial property 
that the standard boost $L(p)$ (and hence
the Wigner rotation $W(\l,p)=L^{-1}(\l p)\l L(p)$) 
depends only on the $4$-velocity
$\p=p/\sqrt{\sm}$, and not on $p$ and therewith not on $\sm$,
cf.~$(5.5)$ of \cite{paper1}. 
Since~\eqref{13.1} is valid for 
all $\psi^-\in\Phi_+$, we can write it as a functional equation in
$\Phi^\times_+$: 
\begin{equation}
\label{13.2}
U^\times(\l,0)|\amd\>=\sum_{j_3'}D^{j_R}_{j_3' j_3}(W(\l^{-1},p))
|\bs{\l}^{-1}\bs{\p}j_3'[\sm_{R} j_{R}]^-\>\,.
\end{equation}
Combining~\eqref{13} and~\eqref{13.2}, the transformation of $|\amd\>$
under $(\l,x)\in{\P}_+$ is given by
\begin{eqnarray}
\lefteqn{U_+^\times(\l,x)|\amd\>}\nonumber\\
& &=e^{-i\gamma\sqrt{\sm_{R}}(t-\bs{x}.\bs{v})}
\sum_{j_3'}D^{j_R}_{j_3' j_3}(W(\l^{-1},p))|
\bs{\l}^{-1}\bs{\p}j_3'[\sm_{R}  j_{R}]^-\>\nonumber\\ 
&&\qquad \qquad \qquad \qquad\text{ only for }\  x^2\geq0,\ t\geq0
\label{13.3}
\end{eqnarray}

The transformation formula \eqref{13.3} of the Gamow kets $|\bs{\p}j_3[\sm_Rj]^-\>\in\Phi_+^\times([\sm_R,j])$
(together with the formula \eqref{13.4} below for the $|\bs{\p}j_3[\sm_R^*j]^-\>\in 
\Phi_-^\times([\sm_R^*,j])$) is the main result of this paper. To appreciate this
transformation formula, we compare it with the unitary representation operator $U^\dagger(\Lambda,x)$
of the Poincar\'e group
\begin{equation}
\nonumber
{\P}=\{ (\Lambda,x)|\Lambda\in\overline{SO(3,1)}, {\text{det}}\Lambda = +1, \Lambda^0_{\hphantom{0}0}
\geq 1, x\in \mathbb{R}_{1,4} \}\,.
\end{equation}
The action of the unitary operator $U^\dagger_{[m^2,j]}(\Lambda,x)=U^\dagger(\Lambda,x)
=U((\Lambda,x)^{-1}) = U(\Lambda^{-1},-\Lambda^{-1}x)$ in the irreducible representation
space ${\cal H}(m^2,j)$ on the momentum basis vectors is written as \cite{joos,weinberg}:
\begin{equation}
\label{x3.7}
{U}^\dagger(\Lambda,x)|\bs{\p},j_3\> = e^{-ip.x}\sum_{j_3'}
D^j_{j_3j_3'}(W(\Lambda^{-1},\p))|\Lambda^{-1}\bs{\p},j_3'\>\,;\,
-\infty < t < \infty\,,
\end{equation}
where $e^{-ipx} = e^{-i\gamma m (t-\bs{v}.\bs{x})}$ and $W(\Lambda^{-1},\p)=L^{-1}(\Lambda^{-1}\p)
\Lambda^{-1}L(\p)$ is the Wigner rotation. The boost $L(\p)$ is given by
\begin{equation}
\label{x3.8}
L^\mu_{\hphantom{\mu}\nu}=
\left(
\begin{array}{cc}
\frac{p^0}{m}&-\frac{p_n}{m}\\
\frac{p^m}{m}&\delta^m_n\!-\!\frac{\frac{p^m}{m}
\frac{p_n}{m}}{1+\frac{p^0}{m}}
\end{array}
\right)\,,
\end{equation}
and acts on the momentum $p^\mu$ (and similarly on the $4$-velocity $\p^\mu=p^\mu/m$)
in the following way:
\begin{equation}
\label{x3.9}
L^{-1}(\p)^\mu_{\hphantom{\mu}\nu}p^\nu=\left(\begin{array}{c}m\\0\\0\\0\end{array}\right)\,.
\end{equation}
Formally, \eqref{x3.7} and \eqref{13.3} look the same.
One just replaces the real mass $\sqrt{\sm}=m$ of the Wigner representation \eqref{x3.7}
by the complex value $\sqrt{\sm}=\sqrt{\sm_R}$. However, \eqref{13.3}
is valid~\footnote{For the other $(\Lambda,x)\notin{\P}_{+}$, the operator
$U^\times(\Lambda,x)$ is not defined, because $U(\l,x)$ is not a continuous
(bounded) operator in $\Phi_+$ and would lead to infinities.} only for
$(\l,x)\in{\P}_{+}=\{(\l,x)\in{\P}, x^2=t^2-\bs{x}^2\geq 0\,,\, t\geq 0\}$.
 Further, whereas \eqref{x3.7}
holds only for real values $m^2$, \eqref{13.3} holds for any complex value $\sm=\sm_R$ 
of the lower complex half-plane, $\sm_R\in{\mathbb C}_-$. In the physical application,
we choose for ${\mathbb C}_-$ the lower half of the second sheet of the Riemann
surface for the $S$-matrix (cf. Figure $2$ of \cite{paper1}) 
and in particular for $\sm_R$ the positions of the
resonance poles on the second sheet of the $S$-matrix. But the formula \eqref{13.1} 
and therewith \eqref{13.3} holds
for any $\sm_R\in{\mathbb C}_-$ as long as $\psi^-\in\Phi_+$. 
The Lippmann-Schwinger kets with the
transformation property \eqref{p10-} are the limiting case.

When we do the contour deformation in \cite{paper1}, going from $(5.14)$ of \cite{paper1}
to $(5.15)$ and $(5.25)$, the real values of $\sm$ in $(5.14)$ are changed to complex
values. The values of $\bs{\p}$ could also have been changed in this process. We decide not
to do this but keep the values of $\bs{\p}$ fixed in the analytic continuation
of the wavefunctions $\<\psi^-|\am\>$ from the physical values $\sm_0\leq \sm<\infty$ 
(second sheet upper rim) 
in $(5.14)$ to complex values of $\sm$. This is possible because the boost
$L(\p)$ and therewith $W(\l,\p)$ depends upon $\p$, not upon the momentum $p=\sqrt{\sm}\p$.
It is this property that allows us to construct the representations $[\sm_R,j]$ by analytic
continuation. The momenta then become ``minimally'' complex, meaning that the momentum
$p$ is given as the product of the complex invariant mass $\sqrt{\sm}$ with the real $4$-velocity
vector $\p$ $(\p_\mu\p^\mu=1)$, $p=\sqrt{\sm}\p$. In these ``minimally complex
representations'', $[\sm_R,j]$, the homogeneous Lorentz transformations $(\l,0)$ are 
represented unitarily as for the unitary representation $[m^2,j]$ of the group $\P$.

To summarize, the semi-group representations
$[\sm_R,j]$ of causal Poincar\'e transformations $\P_+$ are characterized by:
\begin{enumerate}
\item spin (parity) $j$ given by the $j^{th}$ partial wave amplitude in which the
resonance occurs:
\begin{equation}
\nonumber
a_j(\sm) = a_j^{BW}(\sm) + B(\sm)\,.
\end{equation}
\end{enumerate}
It represents the spin $j$ of the resonance.
\begin{enumerate}
\item[2.] the complex mass squared $\sm_R$ (with $\text{Im}\,\sm_R < 0$) given by the resonance
pole position on the second sheet of $S_j(\sm)$ or $a_j^{BW}(\sm)$.
\end{enumerate}
It is connected to mass $M_R$ and width $\Gamma_R$ of the resonance by $\sm_R=M_R-i\Gamma_R/2$.
\begin{enumerate}
\item[3.] minimally complex momenta, $\bs{p}=\sqrt{\sm_R}\bs{\p}$ where $\{\bs{\p}\}={\mathbb R}^{3}$.
\end{enumerate}
The restriction to ``minimally complex'' representations of the Poincar\'e transformations
is necessary, because we need to assure that the $4$-velocity $\p$ is real, since the boost
(rotation-free Lorentz transformation from rest to the $4$-velocity $\p$ or the
three-velocity $\bs{v}=\bs{\p}/\gamma$, $\gamma=1/\sqrt{1-\bs{v}^2}$) is a function
of a real parameter $\p$. The condition $3$ also assures that the restriction of the
representation $[\sm_R,j]$ to the homogeneous Lorentz subgroup is the same unitary
representation as occurs in Wigner's unitary representation for stable particles
$[m^2,j]$. In this way, Wigner's representations $[m^2,j]$ for stable particles
are something like a limiting case of the semigroup representations $[\sm_R=(M_R-i\Gamma_R/2)^2,j]$
for quasistable particles, and the concept of spin $j$, which labels the partial wave
in which the resonance occurs, retains its meaning.

In the same way one derives the transformation property of the Gamow
kets $|\bs{\hat{p}}j_3[{\sm}_R^*j_R]^+\>$ associated with the
resonance pole in the upper half energy plane at
$\sm_R^*=(M+i\Gamma/2)^2$ under $(\Lambda,x)\in{\cal
P}_-$:
\begin{eqnarray}
\lefteqn{U_-^\times(\l,x)|\bs{\hat{p}}j_3[{\sm}_R^*j_R]^+\>}\nonumber\\
& &=e^{-i\gamma\sqrt{\sm_{R}^*}(t-\bs{x}.\bs{v})}  
\sum_{j_3'}D^{j_R}_{j_3' j_3}(W(\l^{-1},p))|
\bs{\l}^{-1}\bs{\p}j_3'[\sm_{R}^*j_{R}]^+\>\,\nonumber\\
&&\qquad \qquad \qquad \qquad\text{ only for}\  x^2\geq0,\
t\leq0\label{13.4} 
\end{eqnarray}
All that has been said above about the representations $[\sm_R,j]$ holds
also for the $[\sm_R^*,j]$, except that here $\text{Im}\,\sm_R^* > 0$ and
these are transformations in the backward light cone.

To emphasize the difference between \eqref{13.3} and \eqref{13.4} we
have labeled the operators $U^\times(\Lambda,x)$ by $\pm$, 
characterizing the operators $U_\pm^\times(\Lambda,x)$ by the same
label by which we characterize the spaces $\Phi_\pm^\times$\footnote{Note that we label
the operators by the same subscript as the spaces which they act
in. Thus, $U_\pm$ act in $\Phi_\pm$, and $U^\times_\pm$ act in
$\Phi_\pm^\times$. But since we use for vectors the standard
physicists' notation $|\bs{\p}j_3[\sm j]^\mp\>\in\Phi^\times_\pm$,
$\psi^-\in\Phi_+,\ \phi^+\in\Phi_-$, the operators $U_\pm^\times$ act
on the vectors $|\bs{\p}j_3[\sm j]^\mp\>$, and $U_+$ acts on $\psi^-$,
$U_-$ on $\phi^+$. We often omit the labels on the operators when the
labels of the vectors imply that they act on a specific space.}. 
$U^\times_\pm$ is, for all $(\Lambda,x)$ for which
it is defined and continuous (in $\Phi^\times_\pm$), the uniquely
defined extension of the same operator $U^\dagger(\Lambda,x)$ in
$\H$. $U^\times_+(\Lambda,x)$ is 
defined for all $(\Lambda,x)\in\P_+$
\begin{equation}
\label{3.8}
U^\times_+(\Lambda,x)\supset U^\dagger(\Lambda,x)\ {\rm in}\
\Phi_+^\times\ {\rm for}\ (\Lambda,x)\in\P_+
\end{equation}
and $U_-^\times(\Lambda,x)$ is defined for all $(\Lambda,x)\in{\cal
P}_-$ 
\begin{equation}
\label{3.9} 
U^\times_-(\Lambda,x)\supset U^\dagger(\Lambda,x)\ {\rm in}\
\Phi_-^\times\ {\rm for}\ (\Lambda,x)\in\P_-
\end{equation} 

Thus we have two different operators $U^\times_\pm$ in two different
spaces $\Phi_\pm^\times$ (defined as the conjugate operators of
$U_\pm=U(\Lambda,x)|_{\Phi_\pm}$, where $U_\pm^\times$ is the
restriction of the unitary group operators $U(\Lambda, x)$)
representing two different subsemigroups $\P_\pm$ of $\P$. 

At this
stage we have no physical interpretation for the operators 
\begin{equation}
U^\times_-(\Lambda,x),\quad U_-(\Lambda,x)\ \text{in the spaces}\
\Phi_-^\times,\ \Phi_-\label{3.10}
\end{equation}
They would represent the semigroup transformations into the
backward light cone. It may be possible that one can find a physical
interpretation 
for them when one takes  
$C$, $P$ and $T$ into consideration. Without that we shall see in the
following section that the semigroup transformations \eqref{3.10} 
would violate the causality conditions for the probabilities \eqref{4.5b}
and may therefore be of no further relevance. 

The results of Sections 2 and 3 have been derived as a mathematical
consequence of the new hypothesis \eqref{pr1} and
\eqref{pr2}. However, even if one does not want to make this
hypothesis but just wants to use the Lippmann-Schwinger kets with an
infinitesimal imaginary part of the energy $p^0$ (or of the invariant
mass $\sqrt{\sm}$ or $\sm$ which has the same effect as long as it is
infinitesimal) one cannot justify the unitary group transformation law
\eqref{v25} for all $x$, $-\infty<x^\mu<\infty$. The transformation
formula \eqref{v25} for the whole Poincar\'e group $\P$ can only
be justified for kets $|\bs\p j_3
[\sm,j]\>\in\Phi^\times\subset\Phi^\times_\pm$, where $\Phi$ is the
space of footnote 1. 
For the
Lippmann-Schwinger kets
$|\bs\p j_3[\sm,j]^-\>$ $=$ \mbox{$|\bs\p j_3 [\sm\mp i\epsilon, j]\>$},  
that require analytic extension into the
complex energy semiplanes, even if the analytic extension is 
only on an infinitesimal
strip below or above the real $\sm$-axis, the unitary representation
\eqref{v25} of the whole Poincar\'e group $\P$ cannot be mathematically
justified.  

We do not know whether the semigroup
transformation laws \eqref{p9minus}, \eqref{p10-} and \eqref{13.3} for
the semigroup $\P_+$ and the semigroup transformation laws
\eqref{p9plus}, \eqref{p10+} and \eqref{13.4} for the semigroup $\P_-$
are less restrictive than our hypothesis \eqref{pr1} and
\eqref{pr2}, in which case our hypothesis $(\ref{pr}_{\mp})$
would be a stronger assumption than we need to obtain a
semigroup. But since we have to use $(\ref{pr}_{\mp})$
 anyway to relate the Breit-Wigner energy distribution to
the Gamow ket \eqref{rgv}, there is little purpose to look for less
restrictive conditions than the Hardy property $(\ref{pr}_{\mp})$,  i.e., analyticity in
the whole semiplane subject to some limits on the growth at infinity
(cf.~Appendix A).  

\section{Physical Interpretation of the Poincar\'e Semigroup
Transformations}\label{sec4} 

 We now want to compare the results of \eqref{13.3} for the semigroup
$\P_+$ with the experimental situation and set aside the semigroup
$\P_-$ (transformations into the backward light cone). The justification
for this will come forth in the process of our discussion. 

The correspondence between theory and experiment is given by the Born
probabilities. The probability for an observable
$|\psi^-(t)\>\<\psi^-(t)|$ in a state $\phi^+$ is given in quantum
theory by 
\begin{subequations}
\label{4.1}
\begin{equation}
\label{4.1a}\tag{\ref{4.1}a}
P(t)=|\<\psi^-(t)|\phi^+\>|^2=|\<\psi^-|\phi^+(t)\>|^2
\end{equation}
which is measured in 
the experiment by 
\begin{equation}
\label{4.1b}\tag{\ref{4.1}b}
P_{\rm exp}(t)=\frac{N_\psi(t)}{N}=\text{ratio of detector counts for 
out-particles described by $\psi$}  
\end{equation}
\end{subequations}
We shall use this probability interpretation of quantum mechanics not
only for states $\phi^+$ but also for 
generalized state vectors $F^-\in\Phi_+^\times$. This has become
standard for the kets with real eigenvalues \eqref{2.15.5} where
$|\<\psi^-|\bs\p j_3 [\sm,j]^-\>|^2$ represents the probability density
for the center of mass energy $\sqrt{\sm}$ in the out-state
$\psi^-$. We shall apply this probability hypothesis also to the Gamow
ket $F^-=|\bs\p j_3 [\sm_R,j]^-\>\in\Phi_+^\times$ of 
\eqref{13.3}. The generalized probability amplitude
\begin{equation}
\<\psi^-|F^-\>=\<\psi^-|\bs\p j_3 [\sm_R,j]^-\>\label{4.1c}
\end{equation}
then represents the probability to detect the
decay products $\psi^-$ in the generalized state
$F^-\in\Phi_+^\times$. Since $\phi^+\in\Phi_-$ in \eqref{4.1a} is
according 
to \eqref{pr1} and \eqref{pr2} also an element of $\Phi_+^\times$,
$\phi^+\in\Phi_+^\times$, the standard probability interpretation \eqref{4.1a}
is just a special case of the probability interpretation for
\eqref{4.1c}. If one takes in place of the Gamow ket
$F^-=|\bs{\p}j_3[\sm_R j_R]^-\>$ the
transformed Gamow ket $U^\times(I,x)|\bs{\p}j_3[\sm_R j_R]^-\>$ 
given by \eqref{13.3} --choosing $\Lambda=1$-- then one obtains the
probability amplitude to detect the decay 
products $\psi^-$ in the evolved Gamow state as 
\begin{equation}
\label{4.2S} 
\<\psi^-|U_+^\times(I,x)|\bs\p j_3
[\sm_R,j_R]^-\>=e^{-i\gamma\sqrt{\sm_{R}}(t-\bs{x}.\bs{v})}\<\psi^-|\bs\p
j_3 [\sm_R,j_R]^-\>
\end{equation}
This evolution of the state is, according to \eqref{13.3}, into the
forward light cone only, 
\begin{equation}
\label{4.2a}
x^2\geq0,\ t\geq0
\end{equation}
Equivalently, \eqref{4.2S} also represents 
the probability amplitude to detect the unevolved
Gamow state with an observable
\begin{equation}
|\psi^-(x)\>\<\psi^-(x)|=
U_+(I,x)|\psi^-\>\<\psi^-|U_+^\times(I,x)\label{4.3}
\end{equation}
which  has been translated from $|\psi^-\>\<\psi^-|$ into the forward
light cone $x^2\geq0,\ t\geq0$
\begin{equation}
\<\psi^-(x)|\bs\p j_3
[\sm_R,j_R]^-\>=e^{-i\gamma\sqrt{\sm_R}(t-\bs{x}.\bs{v})}\<\psi^-|\bs\p
j_3[\sm_R,j_R]^-\>.\label{4.4}
\end{equation} 
The l.h.s of \eqref{4.2S}  and \eqref{4.4} are the same quantity
looked at from the 
Schroedinger and Heisenberg picture, respectively. 
In either case the spacetime translations of the probability
(amplitude) is only into the forward light cone $x^2\geq0,\
t\geq0$. This light cone condition we write in two parts:
\begin{subequations}
\label{4.5}
\begin{equation}
\label{4.5b}
\tag{\ref{4.5}a}
t\geq0
\end{equation}
and
\begin{equation}
\label{4.5a}
\tag{\ref{4.5}b}
t^2\geq\bs{x}^2\equiv r^2/c^2
\end{equation}
\end{subequations}
These two parts \eqref{4.5b} and \eqref{4.5a} express two versions of
causality.  

We first consider the decaying state in the rest frame. Then,
$\bs{v}=\frac{\bs{\p}}{\gamma}=0$ and the Poincar\'e transformation
\begin{equation}
\<U_+(I,x)\psi^-|\bs{0}j_3[\sm_Rj_R]^-\>
=e^{-i\sqrt{\sm_R}t}\<\psi^-|\bs{0}j_3[\sm_Rj_R]^-\>,\ t\geq0\label{4.5c}
\end{equation}
is the time evolution in the rest frame. This time evolution starts at
the mathematical time of the semigroup $t=0$. This introduces a new
concept: the semigroup time $t=0$ 
is the time at which the decaying state $|\bs{0}j_3[\sm_Rj_R]^-\>$ has
been created and the registration of the decay products (described by
the projector on the out-state vector $\psi^-(x)$) can be done. This
semigroup time $t=0$ can be an arbitrary point in the time of our lives, and
we call this arbitrary time at which the decaying particle has been
produced, the time $t_0$, This arbitrary time $t_0>-\infty$ has
been identified with the mathematical semigroup time $t=0$. (The
requirement $t_0>-\infty$ assures that it is not a time symmetric unitary group
evolution). The time $t_0$ is a new concept which has been introduced
by the semigroup and which has  no
place in the standard quantum theory because the time evolution in the
Hilbert space is given by a unitary group $e^{iHt}\psi,\
-\infty<t<\infty$ (the solution of the Heisenberg equation in the
Hilbert space)\footnote{This follows from the Stone-von Neumann
theorem.}. 
The condition \eqref{4.5b} then says that a state
needs to be prepared first, at $t=t_0\,(=0)$ before one has a probability
proportional to the modulus square of the amplitude \eqref{4.2S}=\eqref{4.4}.
 
The condition \eqref{4.5a}, $t-t_0\geq r/c$, says that the probabilities
{\em cannot} propagate with a velocity $r/{(t-t_0)}$ faster than the
velocity $c$.

The condition \eqref{4.5a} is fulfilled for both semigroup
transformations $\P_+$ and $\P_-$ but not for all Poincar\'e
transformations $\P$. The condition \eqref{4.5b} is fulfilled for
transformations of $\P_+$ (forward light cone) only, not for the
semigroup $\P_-$. (This is the reason we have set aside the semigroup
transformations \eqref{p9plus}, \eqref{p10+} \eqref{13.4}, at least for
the time being.)

\begin{figure}[p]
\scalebox{.9}{\includegraphics{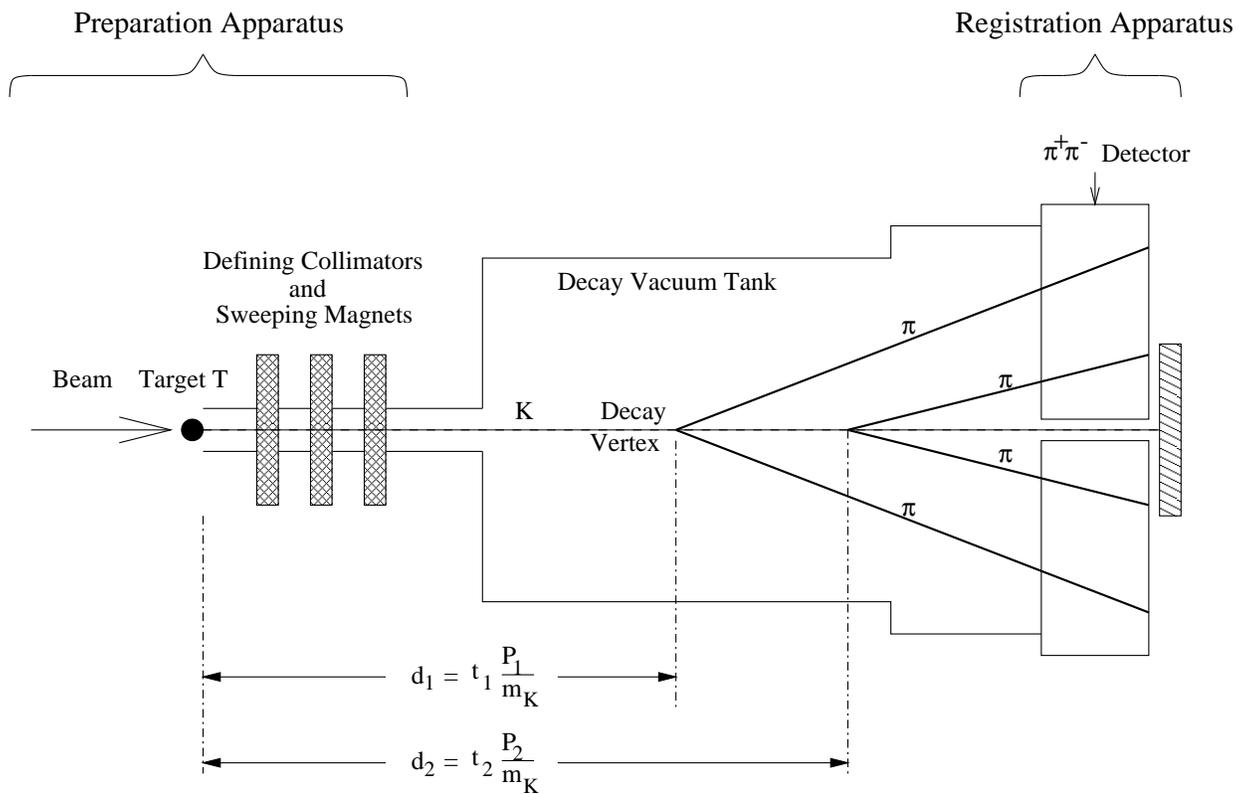}}
\caption{Simplified diagram of the typical neutral $K$-meson decay experiment.}
\label{kmeson_fig}
\end{figure}

The forward-light-cone condition \eqref{4.5b}, \eqref{4.5a} 
expresses the intuitive notion of causality, that a state must be
prepared first before an observable can be measured in it and that the
probability for the observable in a prepared state cannot propagate
faster than with the velocity of light. This intuitively evident,
causality condition is here obtained as a consequence of the new
hypothesis (\ref{pr}$_\mp$) and is not fulfilled by unitary group
representations in Hilbert space. To go into more detail, 
we shall use for our discussions of the 
correspondence between theory and experiment the decay of the neutral
Kaons (as occurs e.g. in the reaction $\pi^-p\rightarrow\Lambda K^\circ$,
$K_S^\circ\rightarrow\pi^+\pi^-$, cf. also Figure $1$ of \cite{paper1}) for
which there exists a series of famous experiments~\cite{?,ref?}. A
simplified schematic diagram of these experiments is given in Figure \ref{kmeson_fig}. 

A $K^\circ$ is produced with a time scale of $10^{-23}$~s by
strong interaction and it decays by weak interaction with a time scale
of $10^{-10}$~s, which is roughly the lifetime of the $K_S^\circ$,
$\tau_{K_s}$. This defines very accurately the time $t_0$ at which the
preparation of the $K^\circ$-state is completed and the registration
of the decay products can begin (theoretical uncertainty is
$10^{-13}\tau_K$). The $K^\circ$-state is created instantly at the
baryon target $T$ (the baryon $p$ is excited from the ground state
(proton) into the $\Lambda$ state, with which we are no further
concerned), and a beam of $K^\circ$ emerges from $T$.  
We imagine that {\em single} Kaons, created at a
collection of initial times $t_0^{(n)}$, are moving into the forward
direction $\bs{x}=(0,0,z)$. 
Each $t_0^{(n)}$ at which the $n$-th Kaon is created is
identified with the same mathematical
semigroup time $t_0=0$ of the transformation formula \eqref{13},
\eqref{13.3}, etc. 

One selects $K^\circ$'s that have a fairly well defined momentum, and
we want to discuss first the case 
that it is described by a Gamow vector $|\amd\>$
with $\sm_R=(M_S-i\Gamma_S/2)^2,\ \frac{\Gamma_S}{M_S}=10^{-14}$, 
and with a well defined 4-velocity
$\bs{\hat{p}}$, i.e., with a momentum
$\bs{p}=\sqrt{\sm_R}\bs{\hat{p}}\approx M_S\bs{\hat{p}}$. Whether such
idealized states exist is the analogue of the question whether
plane-wave states of stable particles $|\bs{p}[mj]\>$ exist; certainly
it cannot be tested experimentally because any macroscopic apparatus
can measure the particle momentum $\bs{p}=m\bs{\hat{p}}$ only within a
certain momentum interval around $\bs{p}$. But we are used to working
with Dirac kets and thinking of them as states with precise 
momentum $\bs{p}$ (or 4-velocity $\bs{\p}$). Below in
\eqref{e1} we will discuss continuous superpositions with sharply peaked
4-velocity $\bs{\p}_0$\footnote{ What we shall not consider here is
that the $K^\circ$ state created at the baryon target $T$ is a
superposition of two neutral Kaons and not a
Gamow vector. This will be discussed briefly at the end of the section,
following \eqref{6.7.1.5}.}.


We consider first the idealized Kaon state described by the Gamow ket 
$|\bs{\hat{p}}[\sm_S]^-\>$ with
$\sqrt{\sm_S}={M_S-i\Gamma_S/2}$. Since $\Gamma_S\ll M_S$, $\Gamma_S$
can be safely neglected when the velocity $\bs{\hat{p}}$ is
experimentally determined as
$\bs{\p}=\bs{p}/M_S$ $=$ \mbox{$({\bs{{p}}}_{\pi^+}+{\bs{{p}}}_{\pi^-})/M_S$}.

Somewhere downstream in Figure \ref{kmeson_fig} at a distance $z=d_1, d_2,\cdots, d_n,\cdots$
from $T$, we  ``see'' a decay vertex for $\pi^+\pi^-$. A detector
(registration apparatus) has been built such that it counts
$\pi^+\pi^-$ pairs which are coming from the position
$\bs{x}=(0,0,z)$. The observable registered by the detector is the
projection operator
\begin{equation}
\Lambda(x)=|\psi^-(t,\bs{x})\>\<\psi^-(t,\bs{x})|
=|\pi^-\pi^+,t\>\<\pi^+\pi^-,t|\label{kaon1}
\end{equation} 
for those $\pi^-\pi^+$ which originate from the fairly well specified
location $x=(t,\bs{x})=(t,0,0,z)$.

More realistically $\Lambda$ should be a projection operator on a
multidimensional
subspace of $\Phi_+$ describing the decay products $\pi^+\pi^-$
counted by the detector (with finite energy and angle resolution) of
which we consider here the one dimensional subspace described by the
pure out-state vector $\psi^-=|(\pi^+\pi^-)\>\in\Phi_+$,
\eqref{kaon1}. This 
means that its   energy wave function $\<^-\sm|\psi^-\>$ is a smooth
Hardy function $\<\sm^-|\psi^-\>\in{\cal H}^2_+$  which can be
analytically continued into the complex $\sm$-plane.

Before we consider the experiment of Figure \ref{kmeson_fig}, let us discuss
the situation that the ($\pi^+\pi^-$)-detector represented by
$|\psi^-\>\<\psi^-|$ is in the rest frame of the decaying
$K^\circ$. In its rest frame, the decaying $K^\circ$ evolves in time
according to
$e^{-iH^\times\tau}|0,\sm_S^-\>=e^{-i\sqrt{\sm_S}\tau}|0,\sm_S^-\>$,
where $\tau$ is the proper time (in the $K^\circ$ rest frame) and
$\sqrt{\sm_S}=(M_S-i\Gamma_S/2)$. Thus, the probability rate density
for counting the $\pi^+\pi^-$ by the detector in the $K^\circ$ rest
frame is according to \eqref{4.2S} and \eqref{4.4} proportional to
\begin{equation}
|\<\psi^-|e^{-iH^\times\tau}|0,\sm_S^-\>|^2
=|\<e^{iH\tau}\psi^-|0,\sm_S^-\>|^2
=e^{-\Gamma_S\tau}|\<\psi^-|0,\sm_S^-\>|^2\label{exponential}
\end{equation}    
This is the usual exponential dependence upon the time $\tau$ in the
rest frame.

In practice \cite{?,ref?}, 
one does not measure the counting rate by detectors in
the rest system of the $K^\circ$, but one has a $K^\circ$ that moves
with a fairly well defined   momentum $\bs{p}$ into the $z$-direction
(beam direction, cf.~Figure \ref{kmeson_fig}).  One measures the counting rate as a
function of the distance $z$ from the position $T$ at
which the $K^\circ$'s have been produced. The formula for this
distance dependence of the counting rate is usually justified from
relativistic  kinematics of classical particles. Here we want to
derive this formula by relativistic quantum theory from the
transformation property~\eqref{13.3} of the Poincar\'e semigroup and
therewith obtain experimental support of the theoretical 
result~\eqref{13.3} derived from the hypothesis (\ref{pr}$_\mp$).

In the experiment depicted schematically in Figure \ref{kmeson_fig}, one has instead
of the detector $|\psi^-\>\<\psi^-|$ and a decaying $K^\circ$-state
$|0,\sm_S^-\>$ at rest a decaying $K^\circ$-state
$|\bs{\hat{p}},\sm_S^-\>$  with 
momentum\footnote{Note that momentum  $\bs{p}$ and mass $\sqrt{\sm_S}$
of the decaying Gamow  state are both complex in such a way 
that $\bs{\hat{p}}$ and
$\bs{v}=\frac{\bs{\hat{p}}}{\gamma};\
\gamma=\frac{1}{\sqrt{1-v^2}}=\sqrt{1+\bs{\hat{p}}^2}={\hat{p}}^0$ are
real. However, since $\frac{{\rm Im}\sqrt{{\sm}_S}}{M_S}\approx 10^{-14}$, 
Im$\sqrt{\sm_S}$ is negligible compared with $M_S$ when
$\bs{\hat{p}}$ is obtained from $\bs{p}=\bs{p}_{\pi^+}+\bs{p}_{\pi^-}$
(the difference between $p^0$ and $M_S\p^0$ is ($-i0$) which means it
is important for the boundary conditions but not for the quantitative
analysis).}
$\bs{{p}}\approx M_S{\bs{\hat{p}}}$    and the detectors
$|\psi(x)^-\>\<\psi^-(x)|$ scan the whole flight path of the Kaon
along the $z$-axis, $\bs{x}=(0,0,z)$ (in the lab frame). 

The detector that counts the decay event
$K^\circ\rightarrow\pi^-\pi^+$ at $\bs{x}$ is obtained from
the detector $|\psi^-\>\<\psi^-|$ at a reference position
$x_1=(t_1,\bs{x}_1)$ (e.g., counting the decay events
$K^\circ\rightarrow\pi^-\pi^+$  at time $t_1=t_0=0$ directly 
at the target position $\bs{x}_1=\bs{x}_T=0$) by a space-time translation
$U(I,x)=U(I,(t,\bs{x}))$:
\begin{equation}
|\psi(x)^-\>=U(I,(t,\bs{x}))|\psi^-\>,\ x^2\geq0,\ t\geq0\label{kaon2}
\end{equation}
In order that this space-time position $(t,\bs{x})$ of the detector
runs 
with the $K^\circ$ that is moving with velocity $\bs{\p}\approx
\bs{p}/M_S$ 
along the $z$=axis, 
the parameters $(t,\bs{x})$ (of the space-time
translation of the classical apparatus) must fulfill
$\frac{\bs{x}}{t}=\frac{\bs{\hat{p}}}{\gamma}=\bs{v}$, since $\bs{v}$
is the velocity in the lab frame of the particle $K^\circ$.
The parameters of the space-time translation $(t,\bs{x})$
from position $(t_0=0, {\bs{x}}_T=0)$  
at which the decaying particle has been created to position 
$(t,0,0,z)$ at which the decay event is counted
(i.e.,
the distance from $T$ to the decay vertex 
in Figure \ref{kmeson_fig}) is thus given by
\begin{equation}
\label{kaon3}
x=(t,\bs{x})=(t,0,0,z=tv)=(\frac{z}{v},0,0,z)
=(\gamma\tau,0,0,\tau\hat{p})
\end{equation}

To obtain the prediction for the measured counting rate
we have thus to calculate the probability density
amplitude $\<\psi^-(x)|{\bs{\hat{p}}},\sm_S^-\>$ and the probability rate
for a decay event $\pi^+\pi^-$ at $x$
which is proportional
to $|\<\psi^-(x)|\bs{\hat{p}},\sm_S^-\>|^2$. 
Using the transformation formula for the Gamow
kets \eqref{13.3} in the probability 
amplitude 
\begin{equation}
\label{kaon4a}
\tag{\ref{kaon4}a}
\<\psi^-(x)|{\bs{\hat{p}}},\sm_S^-\>
=\<U(I,x)\psi^-(x)|{\bs{\hat{p}}},\sm_S^-\>
=\<\psi^-|U^\times(I,x)|\bs{\hat{p}},\sm_S^-\>,\  t\geq0,\
t\geq\bs{x}.\bs{v}
\end{equation}
we obtain\footnote{The inequalities in \eqref{kaon4a} and
\eqref{kaon4} are the causality conditions \eqref{4.5a} and
\eqref{4.5b}: $t\geq0,\ x^2=t^2-z^2=t^2(1-v^2)\geq0$} from \eqref{4.2S} 

\begin{eqnarray}
\<\psi^-(t=\frac{z}{v},0,0,z)|{\bs{\hat{p}}},\sm_S^-\>
&=&e^{-i\gamma\sqrt{\sm_S}(\frac{z}{v}-zv)}
\<\psi^-|{\bs{\hat{p}}},\sm_S^-\>\nonumber\\
&=&e^{-i(M_S-i\Gamma_S/2)\frac{z}{\gamma v}}
\<\psi^-|{\bs{\hat{p}}},\sm_S^-\>\nonumber\\
&&\text{for}\  t\geq0,\ 1\geq v\label{kaon4}
\end{eqnarray}

Therewith we predict that the probability rate as a function of the
distance $z$ from the target is proportional to
\begin{equation}
|\<\psi^-(x)|{\bs{\hat{p}}},\sm_S^-\>|^2=e^{-\Gamma_S\frac{z}{\gamma
 v}}|\<\psi^-|{\bs{\hat{p}}},\sm_S^-\>|^2
=e^{-\frac{z}{\gamma\beta c}\Gamma_S}
|\<\psi^-|{\bs{\hat{p}}},\sm_S^-\>|^2,\quad t\geq0,\ 1\geq v/c\label{kaon5}
\end{equation}
where we have reverted to the standard units with light speed $c$ and
$\beta=\frac{v}{c},\ \gamma=\frac{1}{\sqrt{1-(v/c)^2}},\
v=\frac{\hat{p}}{\gamma}=\frac{p}{\gamma (M_S-i\Gamma/2)}\approx\frac{p_z}{\gamma M_S}$. 
The momentum 
$p_z$ is measured as the $z$-component of the $\pi^+\pi^-$-system,
$p_z=(p_{\pi^+}+p_{\pi^-})_z$. With \eqref{kaon3}, the exponential in
\eqref{kaon5} is  
\begin{equation}
\label{kaon5'}
\tag{\ref{kaon5}a}
e^{-\Gamma_S\frac{z}{\gamma
v}}=e^{-\Gamma_St/\gamma}=e^{-\Gamma_S\tau},\quad \tau\geq0
\end{equation}
The result \eqref{kaon5} is identical to the 
formula used for fitting 
the $\pi^+\pi^-$ counting rate, e.g., equation (23), (24) of \cite{ref?} for
$K^\circ$ and (1) of \cite{ref??} for the $B^\circ$. It has 
the advantage of fitting the rate as a function of distance making
use of time dilation. But even more important is the fact that
\eqref{kaon5} does not require  
the knowledge of the creation times $t_0^{(n)}$ for each
individual member of the ensemble of $K^\circ$'s. 

We thus have the following situation: An ensemble of $K^\circ$'s is
created at various times $t_0^{(n)}$ in the laboratory (over months
etc of the run of the experiment) at the position $T$ with $z=0$. The
$n$-th $K^\circ$ moves down the beam line during the time interval
$t_n-t_0^{(n)}$ and decays at $t_n$ after it has moved the distance
$z=v(t_n-t_0^{(n)})=\frac{p_z}{\gamma M_S}(t_n-t_0^{(n)})=v\gamma\tau_n = \frac{p_z}{M_S}\tau_n$. The 
ensemble of $K_S^\circ$ created at these different times
$t^{(n)}_0$, which are {\em all} represented by the same semigroup
time $t_0=0$,  
 is described by the (almost) momentum
eigenvector $\phi^G_{\sm_S}(t)=U^\times(I, (t,0,0,t\bs{v}))|\bs\p
\sm_S^-\>$ (or by $\phi^G_{p_0\sm_R}$ below). This generalized state
vector is evolving in spacetime (starting at $(t_0=0,\ \bs{x}_T=0$) 
and as a consequence the probability
rate for the $\pi^+\pi^-$ at $(t,0,0,z)$ changes according to
\eqref{kaon5}. It is this probability rate as a function of
$z=\frac{p}{\gamma M_S}(t-0)$ which is measured by the counting rate
$\frac{\Delta N(t)}{\Delta t}$ (number of decay events $\Delta N(t)$
per time interval $\Delta t$) at the discrete set of points
$d_n=\frac{p}{\gamma M_S}(t_n-t_0^{(n)})
=\frac{p}{\gamma M_S}(t-0)=z$, which is then fitted to \eqref{kaon5} in order
to determine the value of $\Gamma_S$ which according to \eqref{kaon5'}
is the inverse lifetime, $\Gamma_S=\frac{\hbar}{\tau_S}$. 

Thus the Kaon-state vector $\phi^G_{\sm_S}(t), t\geq0$ describes an
ensemble of individual $K^\circ_S$ (with the same momentum $p$) which are
created at quite different times $t_0^{(n)}$. All these times
$t_0^{(n)}$ in the past of the individual
$K_S^\circ\rightarrow\pi^+\pi^-$ events are the initial time $t_0=0$
for the $K^\circ$-state $\phi_{\sm_S}^G(t)$. This time $t_0$ at which
$\phi^G_{\sm_S}$ has been created and after which one can count the
decay products, i.e., the time $t=0$ in the ``life'' of each individual
$K^\circ$, is identified with the mathematical semigroup time
$t=0$. The vector $\phi^G$ does {\em not} represent a bunch (wave packet) of
$K^\circ$'s moving down the beam line together. But it represents 
an ensemble of $K^\circ$'s 
which are created at quite arbitrary times $t_0^{(n)}$ under the same conditions. They have a well
defined lifetime $\tau={\rm average\ of}\ \frac{1}{\gamma}(t_n-t^{(n)}_0)$.
  
In the past, \eqref{kaon5} 
has been justified by applying relativistic kinematics to the
$K^\circ_S$ and treating it as a classical particle. Here we have
derived it from the transformation property of the projection
operator $|\psi^-\>\<\psi^-|$ which represents the 
registration apparatus of the decay products.  
The prediction \eqref{kaon5} also
contains the time asymmetry $t_n-t_0^{(n)}=t>0$, which is also always
tacitly assumed because it is an 
obvious consequence of our feeling for causality (the decay products
can only be counted {\em after} the preparation of each $n$-th $K^\circ$ at
the position ${\bs{x}}_T$ at $t_0^{(n)}$). Here it is also a consequence of \eqref{13.3}.

Since we use the relativistic Gamow vectors
$|{\bs{\hat{p}}},\sm_S^-\>$ defined from the position of an $S$-matrix
pole at $\sm_S=(M_S-i\Gamma_S/2)^2$, we also derive (by~\eqref{kaon4} and~\eqref{kaon5}) that the
inverse of $\Gamma_S$  is
exactly the lifetime $\tau_S$ in the rest frame, 
$\tau_S=\frac{\hbar}{\Gamma_S}$. This is also often 
tacitly assumed but has previously only been justified by the
Weisskopf-Wigner approximation in the non-relativistic case
\cite{weisskopf}. This result, which one can only obtain for the
relativistic Gamow vector with Breit-Wigner energy distribution \eqref{rgv}, is
the reason for which we prefer the parameterization
$\sqrt{\sm_R}=(M_R-i\Gamma_R/2)$, or, the definition $\Gamma_R=-2{\rm
Im}\sqrt{\sm_R}$ over other  definitions, (5.37) of
\cite{paper1}, for the width of the lineshape of a relativistic
resonance \cite{ambiguity,ray}.

We shall now relax the assumption of an exact 4-velocity eigenstate
$|{\bs{\hat{p}}},\sm_S^-\>$ for the $K^\circ$ and start from the
assumption that $K^\circ$ is represented by a resonance state which
has a realistic (not $\delta^3(\bs{\hat{p}}-\bs{\hat{p}}_0)$) 4-velocity
distribution $\phi_{j_3}(\hat{p})$ which however is strongly peaked at
the value $\bs{\hat{p}}_0$. This will lead to results which can be directly
connected to formulas previously given by some heuristic arguments
which also made use of 4-velocity eigenvectors for unstable
relativistic particles~\cite{zwanziger}. 

From \eqref{13.3} it follows that 
the space-time translation of a momentum wave-packet~\eqref{4.10.8}
peaked at $\bs{p}_0$ is given by
\begin{eqnarray}
\label{e1}
U^\times(I,x)\phi^G_{\bs{p}_0\sm_R}
=\sum_{j_3}\int\frac{d^3\bs{\p}}{2\p^0}
e^{-i\gamma\sqrt{\sm_R}(t-\bs{x}.\bs{v})}|\amd\>\phi_{j_3}(\bs{\p})&&\nonumber\\
\text{only for } t\geq 0,\ t\geq\bs{x}.\bs{v}&& 
\end{eqnarray}
This represents the Gamow vector with a 4-velocity distribution
described by the wave function
$\phi_{j_3}(\bs{\hat{p}})=\phi_{j_3}(\gamma{\bs{v}})$ which has been
time and space translated by the 4-vector $(t,\bs{x})$. Therefore the
decay probability amplitude for this Gamow state $\phi^{\rm
G}_{p_0\sm_R}$ is (with $j=0$ for simplicity):
\begin{eqnarray}
\<\psi^-(x)|\phi^{\rm G}_{p_0\sm_R}\>&=&\<\psi^-|U^\times(I,x)|\phi^{\rm
G}_{\bs{p}_0\sm_R}\>\nonumber\\
&=&\int\frac{d^3{\bs{\hat{p}}}}
{2\hat{p}^0}e^{-i\sqrt{\sm_R}\gamma(t-{\bs{x}}{\bs{v}})}
\<\psi^-|{\bs{\hat{p}}},\sm_R^-\>\phi({\bs{\hat{p}}})\label{kaon7}
\end{eqnarray}
Here $t$ and $\bs{x}$ are the time and position at which the detector
counts the decay events $\pi^+\pi^-$, and
${\bs{\hat{p}}}=\frac{\bs{p}}{\sqrt{\sm_R}}$, $\bs{v}$ and
$\gamma=\hat{p}^0$ refer to the prepared state of the $K^0$.

With the assumption that $\phi_{j_3}(\bs{\p})$ is strongly peaked about
$\bs{\p_0}=\frac{\bs{v_0}}{\sqrt{1-\bs{v_0}^{2}}}=\gamma_0{\bs{v_0}}$,  
we can approximate the exponent in~\eqref{e1}
$$
-i\gamma\sqrt{\sm_R}(t-\bs{x}.\bs{v})=
-i\sqrt{\sm_R}\left(\sqrt{1+\bs{\p}^{2}}t-\bs{x}.\bs{\p}\right)
$$ 
by expanding it around
$\bs{\p_0}$ and retaining only the first order terms in $\bs{\p}-\bs{\p_0}$.
Using the first order Taylor expansion of $\sqrt{1+\bs{\p}^{2}}$
\begin{equation}
\label{e2}
\sqrt{1+\bs{\p}^{2}}\approx\sqrt{1+\bs{\p_0}^{2}}
+\frac{\bs{\p_0}.(\bs{\p}-\bs{\p_0})}{\sqrt{1+\bs{\p_0}^{2}}}\,,
\end{equation}
in~\eqref{e1}, we obtain the approximate expression

\begin{equation}
\<\psi(x)^-|\phi^G_{\bs{p}_0\sm_R}\>\approx 
e^{-i{\sqrt{\sm_R}}t/\gamma_0}
\int\frac{d^{3}\p}{2\p^0}
e^{i{\sqrt{\sm_R}}{\bs{\hat{p}}}({\bs{x}}-{\bs{v_0}}t)}
\<\psi^-|{\bs{\hat{p}}},\sm_R^-\>\phi(\bs{\hat{p}})\label{e5}
\end{equation}
where we use $\bs{v_0}=\frac{\bs{\hat{p}_0}}{\gamma_0},\
\gamma_0=\sqrt{1+{\bs{\hat{p}}_0}^2}$.

For
easier interpretation this is written as
\begin{equation}
\<\psi^-(x)|\phi^{\rm G}_{\bs{p}_0\sm_R}\>\approx
e^{-i\sqrt{\sm_R}\frac{t}{\gamma_0}}e^{i\sqrt{\sm_R}
({\bs{x}}-{\bs{v_0}}t){\bs{\p_0}}}A({\bs{x}}-{\bs{v_0}}t),
\quad t\geq0,\ t\geq\bs{x}.\bs{v}_0\label{kaon8}
\end{equation}
where, following~\cite{zwanziger}, $A({\bs{x}}-{\bs{v_0}}t)$ is
defined as
\begin{equation}
\label{e6}
A(\bs{x}-\bs{v_0}t)=\int \frac{d^{3}\p}{2\p^0}
e^{i\sqrt{\sm_R}(\bs{\p}-\bs{\p_0})(\bs{x}-\bs{v_0}t)}
\<\psi^-|{\bs{\hat{p}}},\sm_R^-\>
\phi(\bs{\p})
\end{equation}

The reason for defining $A$ in this way is that for a sharp 
4-velocity distribution of just one value
$\bs{\hat{p}}_0=\gamma_0\bs{v_0}$ defined by
\begin{equation}
\phi(\bs{\hat{p}})=2{\hat{p}}^0\delta(\bs{\hat{p}}-\bs{\hat{p}_0})
\label{e6.1} 
\end{equation}
one obtains for \eqref{e6}
\begin{equation}
A(\bs{x}-{\bs{v_0}}t)=\<\psi^-|\bs{\hat{p}_0}\sm_R^-\>
\end{equation}
Inserting this into \eqref{kaon8}, the probability density amplitude
for this Gamow state $\phi_{p_0\sm_R}^{G}$ in the limit of sharp 4-velocity
$\bs{\hat{p}}_0$ is 
\begin{equation}
\<\psi^-(x)|\phi^{\rm G}_{\bs{\p}_0\sm_R}\>\approx
e^{-i\sqrt{\sm_R}\gamma_0(t-{\bs{x}}.{\bs{v_0}})}\<\psi^-|\bs{\hat{p}}_0\sm_R^-\>
,\quad t\geq0, \quad t\geq\bs{x}.\bs{v}
\label{kaon9}
\end{equation}
Here we have not put any condition on the position $\bs{x}$ around
which the $\pi^+\pi^-$ events are counted. If we now set the detector
such that $\pi^+\pi^-$ events are counted at the position downstream
at ${\bs{x}}={\bs{v_0}}t=(0,0,z)$, then we obtain from \eqref{kaon9},
\begin{equation}
\<\psi^-(t=\frac{z}{v_0},0,0,z)|\phi^{\rm G}_{\bs{p}_0\sm_R}\>
\approx
e^{-i(M_R-i\Gamma_R/2)\frac{z}{v_0\gamma_0}}A(\bs{x}-\bs{v}_0t)\quad 
t\geq0
\label{kaon10.1}
\end{equation}
so that the counting rate is predicted again to be proportional to
\begin{equation}
|\<\psi^-(t=\frac{z}{v_0},0,0,z)|\phi^{\rm G}_{\bs{p}_0\sm_R}\>|^2\approx
e^{-\Gamma_R\frac{z}{v_0\gamma_0}}A(\bs{x}-\bs{v}_0t)\quad 
t\geq0 
\label{kaon10}
\end{equation}
which agrees with~\eqref{kaon5} for $\bs{\hat{p}}_0=\bs{\hat{p}}$ 
as it should for the sharp velocity distribution \eqref{e6.1}.
$A(\bs{x}-\bs{v_0}t)$ thus describes the deviation of the $K^\circ$
beam from a sharp momentum beam.

The expressions~\eqref{kaon8} and~\eqref{e6} --which agree
with~\cite{zwanziger} except that here causality \eqref{4.5} is
also a result-- represents a wave packet 
traveling   with velocity $\bs{v_0}$ and simultaneously decaying 
exponentially with a lifetime
\mbox{$\tau_R=1/\Gamma_R$}. Thus the
probability amplitude for the decay events
$K^\circ\rightarrow\pi^+\pi^-$ is a wave packet that travels with
velocity ${\bs{v_0}}={\bs{\hat{p}_0}}(1+{\bs{\hat{p}_0}})^{-1/2}$, where
$\bs{\hat{p}_0}$ is the central value of the sharply peaked 4-velocity
distribution in the prepared $K^\circ$-state, and decays in
time. This, however, does not mean that a wave packet of
$K^\circ$-mesons is traveling down the $z$-direction because the time
$t$ is the time interval from the creation of the $n$-th  
$K^\circ$ at the
time $t_0^{(n)}$ and this time $t_0^{(n)}$ is a different time by the clocks
in the lab for each single $K^\circ$ decay event. The state $|\phi^{\rm
G}_{\sm_R}\>\<\phi^{\rm G}_{\sm_R}|$ describes an ensemble of single microphysical
decaying systems $K^\circ$ each of which has been produced by the macroscopic
preparation apparatus and a quantum scattering process at different
times $t_0^{(n)}$ in the lab. All of these $t_0^{(n)}$ are mathematically
represented by the semigroup time $t=0$ which represents the creation
time in the life of
each $K^\circ$. Each event (labeled by $n$) 
counted by the detector at the position $z=d_n=v_0(t_n-t_0^{(n)})$ 
is the result of the decay of such a single microsystem that was
created at  $t_0^{(n)}$ and traveled the distance $d_n$ (the different 
$t_0^{(n)}$ can be days apart).

For a detector counting the events at $d_n$, the ensemble of decaying
$K^\circ$-mesons is not a wave packet traveling in the $z$-direction
but it is an ensemble of individual decay events of Kaons which were 
created at the times $t_0^{(n)}$. And each individual time
$t_0^{(n)}$ is equal to the semigroup time $t=0$ of the causal
Poincar\'e transformations \eqref{13.3} and \eqref{p10-}.

To conclude this section, we wish to emphasize that neither the
momentum wave-packets \eqref{4.10.8} of \eqref{e1} nor the 
eigenkets $|\bs{\p}[\sm_R]^-\>$ of \eqref{13.3} represent precisely
the apparatus 
prepared Kaon state vector (besides the fact that a realistic prepared
$K^\circ$ state is not a pure state  but a mixture). 
The in-state vector $\phi^+_{K^\circ}\in\Phi_-$ of the
$K^\circ$-beam, prepared by the accelerator and by scattering on the
baryon target $T$, is given by the complex basis vectors expansion
\eqref{n1.22}, \eqref{n1.25}. For the case of a double resonance system, such as
the $K_S$--$K_L$ 
system with 
resonance poles at $\sm_{L/S}=(M_{L/S}-i\Gamma_{L/S}/2)^{2}$ in the
$j_R=0$ partial wave, it is
given according to \eqref{n1.25} and \eqref{4.10.8} by
\begin{eqnarray}
\label{6.7.1.5}
\phi^+_{\bs{p}_0K^\circ}&=&\int\frac{d^3\p}{2\p^0}(|\bs{\p}j_3[\sm_S
j_R]^-\>+|\bs{\p}j_3[\sm_L j_R]^-\>)\phi(\bs\p)
+|{B}\>\\
&=&\phi^G_{\sm_S}+\phi^G_{\sm_L}+|B\>\label{K_SK_L}\nonumber
\end{eqnarray}
where $\phi^G_{\sm_{S,L}}$ is the Gamow vector of $K^\circ_{S,L}$ that
evolves exponentially by the exact Hamiltonian $H=H_0+H_W$ and
$|B\>$ is the background vector  representing
the non-resonant background in the $K^\circ$-production.  The state
vectors $\phi^G_{\sm_{S,L}}$ describe the exponential decay and 
$|B\>$ is an integral over the energy-continuum \cite{bohm.k}.
The 4-velocity wave function $\phi(\bs\p)$ is peaked at a value
$\p_0\approx\frac{p_0}{M_S}\approx\frac{p_0}{M_L}$

In the Weisskopf-Wigner
approximation~\cite{weisskopf}, which amounts to the 
omission of the background integral $|B\>$, 
\eqref{6.7.1.5} reduces to the superposition of $K_S$- and $K_L$-Gamow
vector:
\begin{equation}
\label{6.7.4}
\phi^+_{K^\circ}\approx \phi^{S}_{\sm_S}+\phi^L_{\sm_L}
\end{equation}
This is the approximation that is 
always used for the $K^\circ$-system and $B^0$-system following \cite{lee}. We apply
now the transformation $U^\times(I,x)$ with $x$ given by \eqref{kaon3}
to \eqref{6.7.4} as done in \eqref{kaon10} etc., and obtain  
in place \eqref{kaon10.1} for the
probability density amplitude of the state 
\eqref{6.7.4}:
\begin{eqnarray}
\label{6.7.6}
\psi^-(t=\frac{z}{v_0},0,0,z)|\phi^+_{K^\circ}\>&\approx&\left(
e^{-i\frac{z}{v_0\gamma_0}(M_S-i\Gamma_S/2)}+e^{-i\frac{z}{v_0\gamma_0}
(M_L-i\Gamma_L/2)}\right)A(\bs{x}-\bs{v_0}t)\nonumber\\
&&\qquad\qquad\qquad\qquad\qquad t\geq0\nonumber\\
\end{eqnarray}
where $A$ is defined in~\eqref{e6}. 
This is the standard expansion used in the $K^\circ$-experiments
\cite{?,ref?}.
It is the superposition of two exponentials. 
The time evolution of the background integral $|B\>$ in
\eqref{6.7.1.5} 
is non-exponential and would lead to 
deviations from the exponential law \eqref{6.7.6}. The time dependence
of the background depends upon the 
preparation of the state $\phi^+$ and thus can vary substantially from
experiment to experiment, whereas the time dependence of the Gamow
state $\phi^{G}_{j_R\sm_{R_n}}$ is always exponential with the same
inverse lifetime $\frac{1}{\tau_{R_i}}=\Gamma_{R_i}$. The lifetimes $\tau_{R_i}$ are 
characteristic of the Gamow states (the resonances per se) and do not vary with the
preparation of the in-state $\phi^+$.

The prediction \eqref{6.7.6} without the background term $|B\>$ 
reduces in the rest frame $\bs{\p}=\bs{0}$ to the time evolution 
obtained in the Lee-Oehme-Yang theory~\cite{lee} 
which uses the heuristic assumption of a complex Hamiltonian. 
Here the time evolution is derived from the transformation property of relativistic
Gamow vectors which describe $K_S$ and $K_L$ as decaying states
defined by the $S$-matrix poles at $\sm_{S,L}$ (resonances).
Thus, the Lee-Oehme-Yang theory, which is based
on the Wigner-Weisskopf approximation~\cite{weisskopf}, 
is recovered in \eqref{K_SK_L} from the exact 
complex basis expansion \eqref{6.7.1.5}  
by neglecting
the background term.
The fact that the $|K_S\>$ and $|K_L\>$ states evolve without mixing
between each other as derived in \eqref{6.7.1.5} and  \eqref{6.7.6} for the Hardy space functionals
$\phi^S_{\sm_S}$ and $\phi^L_{\sm_L}$, cannot be obtained if one uses 
non-exponential Hilbert space vectors for $|K_S\>$ and $|K_L\>$. 
In fact, in an exact Hilbert space
theory for the Kaon system, an (unobserved) vacuum regeneration 
phenomenon between
$|K_S\>$ and $|K_L\>$ is unavoidable \cite{khalfin}. 

Comparing \eqref{6.7.1.5} with \eqref{6.7.4}, 
the question arises: under what conditions and to what extent can 
the background term be neglected
and the quasistable states be isolated by the preparation process? 
Even though a theoretical answer
is not available, the accuracy with which the exponential decay law
and \eqref{6.7.6} 
has been observed in some cases (\cite{norman}, \cite{k+}, and in
particular \cite{ref?}) indicates that, at least 
for $\frac{\Gamma}{m}\approx10^{-14}-10^{-8}$,
the resonance state can be isolated 
from its background to a very high degree of accuracy. 

The situation is quite different for
$\frac{\Gamma}{m}\approx10^{-2}-10^{-1}$ (hadron resonances but also
the $Z$-boson) where $\Gamma$ is measured (not as the inverse lifetime 
but) as the width of the relativistic Breit-Wigner which
according to \eqref{rgv} is the energy wave function of 
the Gamow vector $\phi^{G}_{\sm_R,j}$. 
It is well known empirically that in  
the fit of the $j$-th
partial scattering amplitude 
to the cross section data 
one always needs, in addition to the
Breit-Wigner representing the resonance per se, a background amplitude
$B(\sm)$ corresponding (theoretically) to the background vector
$|B\>$. Thus for larger values of $\frac{\Gamma}{m}$, $B(\sm)$
and thus $|B\>$ may not be negligible.  

The initial decay rate $\frac{\Gamma}{\hbar}\equiv{\rm inverse\
lifetime}\equiv\frac{1}{\tau}$, as measured by a fit of the counting rate to 
\eqref{kaon5} or \eqref{kaon10}, 
and the resonance width $\Gamma_R$ of the relativistic Breit-Wigner as
it appears in \eqref{rgv}, are conceptually and observationally different
quantities. The width $\Gamma_R$ is measured by the Breit-Wigner line shape in
the cross section and the inverse lifetime is measured by the decay
rate as a function of time $t$ or distance $z$,
as in \eqref{kaon10} or \eqref{6.7.6}. The theoretical
connection between these two quantities is given by the Gamow vectors
which according to \eqref{rgv} have a Breit-Wigner energy wave
function and the transformation property \eqref{13.3} 
from which one predicts \eqref{kaon5}. This leads to
$\frac{1}{\tau}=\frac{\Gamma_R}{\hbar}$. Without the Poincar\'e semigroup
representations this relationship cannot be established, although it
is well accepted in non-relativistic quantum mechanics. In the relativistic case
one was never quite sure whether it makes sense to speak of a
resonance per se that can be defined unambiguously as an entity
separated from the background and other resonances.
This ambiguity in the
definition of resonance mass and width in the relativistic regime has
recently been debated extensively in the literature \cite{ambiguity}. The
Poincar\'e semigroup representations $[j_R,\sm_R=m_R-i\Gamma_R/2]$
provide an unambiguous definition 
of a relativistic resonance and its width and mass \cite{ray}. 
The relativistic Gamow kets that span
these semigroup representations 
have the additional attractive feature 
that they generalize Wigner's definition of relativistic
stable particles \cite{wigner}.

\section{Summary and Conclusion}
The purpose of the paper was to give a detailed derivation of the
properties of vectors which describe relativistic decaying states and
resonances. In contrast to non-relativistic quantum theory, where one
considered quasistable particles as unique states characterized by two
numbers $E_R$ and $\Gamma_R=\frac{\hbar}{\tau}$ which can appear
either as a resonance described by a Breit-Wigner energy distribution
or as a decaying particle with an exponential time evolution,
relativistic resonances were considered as complicated objects. Most
commonly, they were defined in perturbation theory by the propagator
using the on-the-mass-shell renormalization scheme and had a mass $M$
and an energy dependent width $\Gamma(\sm)$, the definition of which
depended upon the renormalization scheme \cite{ambiguity}. The
definition of a resonance by the pole of the relativistic $S$-matrix
is less arbitrary and characterizes the resonance by a complex pole
position $\sm_R$, which however does not define  the resonance width
$\Gamma_R$ and resonance mass $M_R$ separately. In the Particle Data Table \cite{groom} the relativistic
quasistable particles are classified by a mass $M_R$ and a width $\Gamma$ or by a mass and a lifetime
$\tau$.

The known relativistic quasistable particles fall into two categories:
those for which one can measure the width of their lineshape (given by
a relativistic Breit-Wigner or others) (when
$\frac{\Gamma}{M_R}\sim10^{-1}-10^{-4}$) and those for which one can
measure the lifetime from the (usually) exponential decay rate (when
$\frac{1}{\tau M_R}\lesssim10^{-8}$). The relativistic Breit-Wigner
lineshape is easily obtained from the $S$-matrix (Laurent
expansion). To obtain the exponential probability rate one postulates
heuristic non-relativistic Gamow functions and $N$-dimensional complex
Hamiltonian matrices for the self-adjoint Hamiltonian operator (e.g.,
$N=2$ for the neutral Kaon system \cite{lee}). The questions how this complex
matrix is derived 
from the self-adjoint Hamiltonian 
$H=P_0$ of the
Poincar\'e time translations or whether $\frac{\hbar}{\tau}=\Gamma_R$
are not addressed, except perhaps in analogy to the non-relativistic
case for which one justifies the lifetime-width relation by the
Weisskopf-Wigner approximation \cite{weisskopf}.

In this and the preceding paper \cite{paper1} we have addressed this
question of the nature of a relativistic quasistable 
particle by starting with the most universally accepted
definition of a resonance by the second sheet pole at $\sm=\sm_R$ of
the analytically continued partial $S$-matrix $S_{j_R}(\sm)$, where
$\sm$ is the square of the scattering energy,
$\sm=p_\mu p^\mu=(p_1+p_2)^2=(p_3+p_4)^2$. This definition by itself is
insufficient to derive satisfactory results because the relativistic
in- and out- Lippmann-Schwinger scattering states
\cite{LS} are ill defined and are attributed contradictory
properties. On the one hand they are to fulfill out-going and
incoming boundary conditions expressed by the $\mp i0$ in the
Lippmann-Schwinger equation and on the other hand they are to furnish a
unitary representation of the Poincar\'e group \cite{weinberg}. To
give the two kinds of Lippmann-Schwinger kets with their
infinitesimally imaginary energy $|\sm^-\>=|\sm_{II}-i0^-\>$ and
$\<^+\sm|=\<^+\sm_{II}-i0|$ (i.e., $|\sm^+\>=|\sm+i0^+\>$) a
mathematical meaning --in the same way as one defines the ordinary Dirac
kets as functionals over the Schwartz space-- we postulated that the
in-states $\phi^+$ and out-observables (often called out-``states'')
$\psi^-$ are elements of two different Hardy spaces \eqref{pr1} and
\eqref{pr2}.  This means the energy wave functions
$\<^+\sm|\phi^+\>=\<^+\bs{\p}j_3[\sm j]|\phi^+\>$ and
$\<^-\sm|\psi^-\>=\<^-\bs{\p}j_3[\sm j]|\psi^-\>$ are postulated to be
smooth Hardy functions \eqref{t4}, \eqref{t3} analytic in the lower and
upper half-plane (second sheet of the $S$-matrix),
respectively. ($\<\psi^-|\sm^-\>=\overline{\<^-\sm|\psi^-\>}$ is Hardy
in the lower half plane). 
Because of the $\mp i\epsilon$, the Lippmann-Schwinger kets
$|\bs{\p}j_3[\sm j]^\mp\>$ do not span a unitary representation $[\sm
j]$ of the Poincar\'e group as usually assumed \cite{weinberg}. But if
defined as functionals $|\bs{\p}j_3[\sm j]^\mp\>\in\Phi^\times_\pm$
one can show that they span an irreducible representation $[\sm\mp
i0,j]$ of the Poincar\'e semigroup transformations $\P_\pm$ into the
forward and backward light cone, respectively \eqref{p10-}, \eqref{p10+}.

The immediate consequence of this is causal propagation of
probability. The $S$-matrix element
$(\psi^-,\phi^+)=(\psi^{out},S\phi^{in})$ given by $(5.1)$ $(5.10)$
of \cite{paper1} represents the probability amplitude to detect the
out-observable $\psi^-$ of \eqref{t2} in the prepared in-state $\phi^+$
of \eqref{t1}. The observable $\psi^-$ can by \eqref{p9minus}
only be predicted 
in the forward light cone relative to the prepared state $\phi^+$ in
$(4.25)$ of \cite{paper1}. This is discussed in Section \ref{sec4}, in
particular for the resonance state described by the Gamow vector. 

Within the mathematical setting provided by the hypothesis \eqref{pr} the Gamow kets
$|\bs{\p}j_3[\sm_R j]^-\>$ can be defined as the vectors with an ideal
Breit-Wigner energy wavefunction \eqref{rgv}. They are continuous
antilinear functionals on $\Phi_+$. Moreover, they are generalized
eigenvectors of the momentum operators with complex eigenvalues
\eqref{htimes}. In particular, they have a complex mass
$\sqrt{\sm_R}$. These complex eigenvalues of the self-adjoint momentum
operators are obtained in the same way
as the ordinary eigenvalues of the Dirac kets, only that the Schwartz
rigged Hilbert space has to be replaced by the pair of Hardy rigged Hilbert
spaces \eqref{pr}.  The Gamow kets have been shown in \cite{paper1} to be
associated with a resonance pole in the second sheet of the
relativistic $S$-matrix at $\sm_R=(M_R-i\Gamma_R/2)^2$. From their
transformation property \eqref{13.3} follows that they span a semigroup
representation space of the causal Poincar\'e transformations \eqref{2.11+}
characterized by $[\sm_R j]$, with $j$ representing spin of the resonating
partial wave and, $\sm_R$, the complex pole position. In order to retain as much
similarity as possible with Wigner's unitary representations \eqref{v25}
of the Poincar\'e group for stable particles $[m^2 j]$ and to maintain
the meaning of the spin $j$ of a resonance, only semigroup
representations with minimally complex momentum are considered, which
means that the momentum is given by $\bs{p}=\sqrt{\sm_R}\bs{\p}$,
where the 4-velocities $\p^0=\gamma=\sqrt{1-\bs{v}^2},\
\bs{\p}=\gamma\bs{v}$ are real. The general transformation formula of
the relativistic Gamow kets under causal Poincar\'e transformations
$\P_+$ is given by \eqref{13.3} which looks very similar to Wigner's
unitary Poincar\'e group transformation, but differs by the property
that transformations are allowed only into the forward light
cone. From \eqref{13.3} follows the exponential time evolution
\eqref{exponential} 
for an isolated Gamow state at rest and the exponential decay law
\eqref{kaon5} for a Gamow state moving with constant velocity $v=\beta c$
along the $z$-direction, as it has been used in experiments \cite{ref?,ref??}.

The prepared state $\phi^+$ is in general not given by a
Gamow vector but it is a linear superposition of Gamow vectors for all
the $N$ resonance poles of the $j$-th partial $S$-matrix $S_j(\sm)$
and in addition there is a background vector \eqref{n1.22}, \eqref{n1.25}. Thus
the prepared state is given by the complex basis vector expansion
\eqref{6.7.1.5}, which  is very similar to the heuristic expansion in terms of
eigenstates with complex eigenvalues $(M_i-\Gamma_i/2)$ for a finite
dimensional effective Hamiltonian, like e.g., of the $K^\circ$-system in
\eqref{6.7.4}. 
Due to the  background vector $|B\>$ over the energy continuum
(which is always lost in the Weisskopf-Wigner approximation)
one obtains in general also deviations from the
exponential decay law for a prepared in-state $\phi^+$. This
explains that in spite of the exponential time evolution for the Gamow
state, describing the resonance per se, one may observe deviations
from the exponential law even if there is only one resonance
present. 

The resonance per se is characterized by $(M_R, \Gamma_R)$
and has exponential evolution and according to \eqref{exponential} 
one predicts for the lifetime of a resonance with Breit-Wigner width $\Gamma_R$ that
$\tau_R=1/\Gamma_R$. 
The validity of lifetime $=$ inverse width relation for $\Gamma_R$ and
not for any of the other width definitions, e.g., $(5.37)$ of \cite{paper1},
removes the ambiguity in the definition of resonance mass and width \cite{ambiguity}
for relativistic resonances. It predicts that the width and mass is given
by $M_R$, $\Gamma_R$ and that the lineshape of the resonance per se is given
by the ``exact'' relativistic Breit-Wigner $(5.29)$ of \cite{paper1}. This also
fixes the background amplitude as the difference of the scattering amplitude
and the ``exact'' Breit-Wigner $(5.25)$ of \cite{paper1}.
The parameters $(M_R,\Gamma_R=\frac{1}{\tau_R})$
do not depend upon the
experiment that prepares the state. For different experiments
with different prepared in-states the
background term in the amplitude and in the state vector \eqref{n1.25},
may change from experiment to experiment, and so will
the deviation from the exponential decay law. But the parameters
$(M_R,\Gamma_R)$ for the resonance per se and the
lineshape of the resonance per se 
will not depend upon the experiment. 
Eliminating the background in the analysis of the
decay data for each particular experiment should reveal the
exponential character of decay.

\section*{Acknowledgement}
We gratefully acknowledge support from the Welch Foundation and express
our gratitude and appreciation to the numerous colleagues and friends
who helped us with criticism and advice. After this paper was completed,
we became aware of a remarkable paper by L.~S.~Schulman \cite{schulman},
who already in 1969 classified the Poincar\'e semigroup representations
and identified our ``minimally complex'' representations $[\sm_R,j]$ 
as the representations of physical unstable particles.

\appendix
\section{Mathematical Results}
Here we prove that for any space-time translation by a $4$-vector 
$x$, such that $x^{2}\geq 0$, $t<0$, there exists $\psi^-\in\Phi_+$
such that $U(I,x)\psi^-\notin \Phi_+$. The proof contained 
in Proposition~\ref{theb} uses simple arguments based on the following
results:
\begin{itemize}
\item {\bf Characterization of Hardy class functions on a 
half-plane}\\
The standard definition of $\H_\pm^{2}$ is the following~\cite{hardy}:
\begin{definition}[$\H_\pm^p$ $1\leq p<\infty$]\label{h:1}
A complex function $f(x+iy)$ analytic in the open lower
half complex plane ${\mathbb C}_{-}$ is said to be a Hardy
class function from below of order $p$, $\H_-^p$, if
$f(x+iy)$ is $L^p$-integrable as a function of $x$ for all $y<0$ and
\begin{subequations}
\begin{equation}
\label{h1}
\underset{y<0}{\rm sup}\int_{-\infty}^{\infty}
dx\,\,|f(x+iy)|^{p}<\infty\,.
\end{equation}
Similarly, a complex function $f(x+iy)$ analytic in the open 
upper half complex plane ${\mathbb C}_{+}$ is said to be a Hardy
class function from above of order $p$, $\H_+^p$, if
$f(x+iy)$ is $L^p$-integrable as a function of $x$ for all $y>0$, and
\begin{equation}
\label{h2}
\underset{y>0}{\rm sup}\int_{-\infty}^{\infty}
dx\,\,|f(x+iy)|^{p}<\infty\,.
\end{equation}
\end{subequations}
\end{definition}
In~\cite{winter}, it is shown that for $p=2$ \eqref{h1} is equivalent
to
\begin{subequations}
\label{wintpm}
\begin{equation}
\label{wint-}
\underset{-\pi<\phi<0}{\rm sup}\int_{0}^{\infty}
|f(r e^{i\phi})|^{2}dr<\infty\,,
\end{equation}
and that \eqref{h2} is equivalent to
\begin{equation}
\label{wint+}
\underset{0<\phi<\pi}{\rm sup}\int_{0}^{\infty}
|f(r e^{i\phi})|^{2}dr<\infty\,.
\end{equation}
\end{subequations}
The definition of Hardy class functions given in \eqref{wint-}
and \eqref{wint+} is the one that we shall use 
for proving our result. This is because the transformation
$z\rightarrow \sqrt{z}$ converts a radial path of integration into
another radial path, while it distorts the horizontal paths of integration
in \eqref{h1} and \eqref{h2}. Further, in the following we use the term Hardy (class)
functions for functions obtained by taking the pointwise limit $y\rightarrow 0$
in \eqref{h:1} and \eqref{wintpm}.\\
\item {\bf Characterization of $\Bmp$}\\
\begin{definition}[$\tilde{\cal S}$]\label{stilda}
The space $\tilde{\cal S}$ is defined as the 
space of Schwartz functions that vanish at zero faster than any
polynomial, i.e., the space of $C^{\infty}$ functions for which
$$
\|f\|_{N}=\underset{\sm\in{\mathbb R}}{\rm sup}\,\,
          \underset{n\leq N}{\rm sup}
\left(|\sm|+\frac{1}{|\sm|}\right)^{N}\left|\frac{d^{n}f(\sm)}{d\sm^n}\right|
<\infty\,,\quad N=0,1,2,\cdots\,.
$$
\end{definition}
\begin{definition}[$\widehat{M}({\mathbb R}_{\pm})$]\label{mhat}
The spaces $\widehat{M}({\mathbb R}_{\pm})$ are the spaces 
of Schwartz functions with supports in ${\mathbb R}_{+}=(0,\infty)$
and ${\mathbb R}_{-}=(-\infty,0)$ with vanishing moments
of all orders, i.e.,
$$
\widehat{M}({\mathbb R}_{+})=\left\{ f\in{\cal S}({\mathbb R})/
\text{supp}f \subset (0,\infty)/ \int_{0}^{\infty}x^{n}f(x)dx=0\,,
\quad n=0,1,2,\cdots\right\}\,,
$$
$$
\widehat{M}({\mathbb R}_{-})=\left\{f\in{\cal S}({\mathbb R})/
\text{supp}f\subset (-\infty,0)/\int_{-\infty}^{0}x^{n}f(x)dx=0\,,
\quad n=0,1,2,\cdots \right\}\,.
$$
\end{definition}
Using the Payley-Wiener theorem, it can be shown 
that the Fourier transform is a continuous bijective functional
from $\widehat{M}({\mathbb R}_{\pm})$ onto $\Bmp$:
\begin{equation}
\label{fouriert}
{\cal F}:\,\,\widehat{M}({\mathbb R}_{\pm})\rightarrow
\Bmp\,.
\end{equation}
\end{itemize}
With these two results, our statement will follow from simple considerations.
\begin{lemma}\label{Fpm}
Left $f\in\Bm$. Define $F_{\mp}(z)$ by  
$$F_-(z)=\frac{f(\sqrt{z})}{z^{1/4}}$$
and
$$F_+(z)=\frac{f(-\sqrt{z})}{z^{1/4}}\,,$$
with the branch of $\sqrt{z}$ and $z^{1/4}$ taken as
\begin{equation}
\tag{\ref{branch}}
-\pi\leq {\rm Arg}\,z<\pi\,.
\end{equation}
Then $F_{\mp}\in\Bmp$.
\end{lemma}
\begin{proof}
We note that, since the branch of $\sqrt{z}$ and $z^{1/4}$ is taken 
on the real axis, $F_{-}$ is analytic on ${\mathbb C}_{-}$
and $F_+$ is analytic on ${\mathbb C}_{+}$. The boundary values
on the real axis of $F_\pm$ are given by
\begin{equation}
\label{bo-}
F_{-}(x)=
\begin{cases}
\frac{f(\sqrt{x})}{x^{1/4}}\,,&\quad x\geq 0\\
\frac{f(-i\sqrt{|x|})}{e^{-i\pi/4}|x|^{1/4}}\,,&\quad x\leq 0\,,
\end{cases}
\end{equation}
\begin{equation}
\label{bo+}
F_{+}(x)=
\begin{cases}
\frac{f(-\sqrt{x})}{x^{1/4}}\,,&\quad x\geq 0\\
\frac{f(-i\sqrt{|x|})}{e^{i\pi/4}|x|^{1/4}}\,,&\quad x\leq 0\,.
\end{cases}
\end{equation}
Since $f\in\tilde{\cal S}$, it follows from \eqref{bo-}
and \eqref{bo+} that $F_\pm\in\tilde{\cal S}$.
It remains to show that $F_{\pm}\in\Bm$.

By performing the transformation $\phi\rightarrow \phi/2$
and $r\rightarrow \sqrt{r}$ in \eqref{wint-}, we obtain
\begin{equation}
\nonumber
\underset{-2 \pi<\phi<0}{\rm sup}\int_{0}^{\infty}
|f(\sqrt{r}e^{i\phi/2})|^{2}\frac{dr}{2\sqrt{r}}<\infty
\end{equation}
Thus
\begin{equation}
\label{up}
\underset{-2 \pi<\phi\leq -\pi}{\rm sup}\int_{0}^{\infty}
|f(\sqrt{r}e^{i\phi/2})|^{2}\frac{dr}{\sqrt{r}}<\infty
\end{equation}
\begin{equation}
\label{down}
\underset{-\pi\leq\phi<0}{\rm sup}\int_{0}^{\infty}
|f(\sqrt{r}e^{i\phi/2})|^{2}\frac{dr}{\sqrt{r}}<\infty\,.
\end{equation}
It follows from \eqref{down} that $F_{-}\in\H_{-}^{2}$.
Changing $\phi$ to $\phi-2\pi$ in \eqref{up}, we obtain
\begin{equation}
\label{updown}
\underset{0<\phi\leq \pi}{\rm sup}\int_{0}^{\infty}
|f(-\sqrt{r}e^{i\phi/2})|^{2}\frac{dr}{\sqrt{r}}<\infty\,.
\end{equation}
Hence $F_+\in\H_{+}^{2}$.
\end{proof}
\begin{proposition}\label{theb}
Given $b>0$, there exists $g\in\Bm$ such that
$e^{i\sqrt{\sm}b}g(\sm)\notin \Bm$, where the branch of $\sqrt{\sm}$
is still given by \eqref{branch}.
\end{proposition}
\begin{proof}
Given a function $h$, let $\tau_a h$ be its translation by $a$:
$\tau_a h(x)=h(x+a)$. Let $h\in\widehat{M}({\mathbb R}_{+})$ be such that
${\rm inf}[\text{supp }h]<b$. Then 
$\tau_{b}h\notin\widehat{M}({\mathbb R}_{+})$.
From \eqref{fouriert}, there exists $f\in\Bm$ such that
$h={\cal F}^{-1}(f)$. Then 
$$
\tau_b h(x)=\int_{-\infty}^{\infty}e^{i\sm(b+x)}f(\sm)\frac{d\sm}{\sqrt{2\pi}}
\notin\Bm\,.
$$
Thus, it follows from \eqref{fouriert} that $e^{ib\sm}f(\sm)\notin\Bm$.
But $e^{ibz}f(z)$ is analytic on ${\mathbb C}_{-}$, and 
$e^{ib\sm}f(\sm)\in\tilde{\cal S}$. Therefore, we deduce that
\begin{equation}
\label{div1}
\underset{-\pi<\phi<0}{\rm sup}\int_{0}^{\infty}
|e^{ibre^{i\phi}}f(r e^{i\phi})|^{2}dr=\infty\,.
\end{equation}
Substituting $\phi\rightarrow \phi/2$, $r\rightarrow\sqrt{r}$
in \eqref{div1}, we obtain
\begin{equation}
\label{div2}
\underset{-2\pi<\phi<0}{\rm sup}\int_{0}^{\infty}
|e^{ib\sqrt{r}e^{i\phi/2}}f(\sqrt{r}e^{i\phi/2})|^{2}\frac{dr}{\sqrt{r}}
=\infty\,.
\end{equation}
Thus
\begin{eqnarray}
\lefteqn{\underset{-2\pi<\phi\leq -\pi}{\rm sup}\int_{0}^{\infty}
|e^{ib\sqrt{r}e^{i\phi/2}}f(\sqrt{r}e^{i\phi/2})|^{2}\frac{dr}{\sqrt{r}}}
\nonumber\\
& &+\underset{-\pi\leq\phi<0}{\rm sup}\int_{0}^{\infty}
|e^{ib\sqrt{r}e^{i\phi/2}}f(\sqrt{r}e^{i\phi/2})|^{2}\frac{dr}{\sqrt{r}}
=\infty\,.\label{div2.5}
\end{eqnarray}
Substituting $\phi\rightarrow\phi-2\pi$ in the first term of
\eqref{div2.5} we obtain
\begin{eqnarray}
\nonumber
\lefteqn{\underset{0<\phi\leq \pi}{\rm sup}\int_{0}^{\infty}
|e^{-ib\sqrt{r}e^{i\phi/2}}f(-\sqrt{r}e^{i\phi/2})|^{2}\frac{dr}{\sqrt{r}}}\\
& & + \underset{-\pi\leq\phi<0}{\rm sup}\int_{0}^{\infty}
|e^{ib\sqrt{r}e^{i\phi/2}}f(\sqrt{r}e^{i\phi/2})|^{2}\frac{dr}{\sqrt{r}}
=\infty\,.\label{div3}
\end{eqnarray}
If the first term in \eqref{div3} is $\infty$, then 
$$
e^{-ib\sqrt{z}}F_{+}(z)\notin\H_{+}^{2}\,.$$
Thus, 
$$
e^{ib\sqrt{z}}F_{+}(z^*)^* \notin \H_{-}^{2}\,.
$$
This is because, as it can be easily seen from \eqref{h:1}, a function
$G(z)\in\H_{+}^{2}$ if and only if $G(z^*)^{*}\in\H_{-}^{2}$. If the 
second term in \eqref{div3} is $\infty$, then
$$
e^{ib\sqrt{z}}F_{-}(z)\notin\H_{-}^{2}\,.
$$
Thus, either $F_{+}(z^*)^*$ or $F_-(z)$ proves the proposition.
\end{proof}
Our result in \eqref{5} is proved by the above proposition, since,
given $x$ and ${\bs{\p}}$ such that $t-\bs{x}.\bs{v}<0$, we have
$$
\<U(I,x)\psi^-|\am\>=e^{ib\sqrt{\sm}}\<\psi^-|\am\>$$
with $b=-\gamma(t-\bs{x}.\bs{v})>0$. From the above proposition, there
exists a function in $\sm$, $\<\psi^-|\am\>\in\Bm$, such that
$\<U(I,x)\psi^-|\am\>\notin\Bm$.

\section{Continuity of $U(I,x)$ on $\Phi_\pm$}\label{o}

We prove here the continuity of $U(I,x)$ on $\Phi_\pm$
under the invariance conditions \eqref{2.12}.

The topology on the spaces 
$\Phi_\pm\equiv\tilde{\cal S}\cap\H_{\mp}^{2}|_{{\mathbb R}_{\sm_0}}\otimes
{\cal S}({\mathbb R}^{3})$ with respect to the $\sm$-variable is given
by the countable number of norms
\begin{equation}
\label{o1}
|\psi^-|_{N}\equiv\|\theta^{-1}\psi^{-}\|_{N}
=\underset{\sm\in{\mathbb R}}{\rm sup}\,\,
          \underset{n\leq N}{\rm sup}
\left(|\sm|+\frac{1}{|\sm|}\right)^{N}\left|\frac{d^{n}}{d\sm^{n}}
\<\psi^-|\am\>\right|
\end{equation}
for $\psi^-\in\Phi_+$, and similarly for 
$\phi^{+}\in\Phi_-$ (see Definition~\ref{stilda}).
The $\theta$ in~\eqref{o1} is the restriction bijective 
mapping~\cite{winter,gadella} in $(5.17)$ of \cite{paper1}:
\begin{equation}
\nonumber
\theta~:~\Bmp\rightarrow~\Bmp|_{{\mathbb R}_{\sm_0}}\,,
\end{equation}
We are concerned here with the topology
on $\tilde{\cal S}\cap\H_\mp^{2}|_{{\mathbb R}_{\sm_0}}$
and not on ${\cal S}({\mathbb R}^{3})$ since continuity
of $U(I,x)$ with respect to the $\bs{\p}$ variable is a 
straightforward Schwartz space result once continuity
with respect to $\sm$ is established.
	With the condition in \eqref{2.12}, i.e., $x^{2}\geq 0$, $t\geq 0$, 
we consider
\begin{eqnarray}
\lefteqn{
\!\!\!\!\!|U(I,x)\psi^{-}|_{N}=
\underset{\sm\in{\mathbb R}}{\rm sup}\,\,
          \underset{n\leq N}{\rm sup}\left(|\sm|+\frac{1}{|\sm|}\right)^{N}
\left|\frac{d^{n}}{d\sm^{n}}\<U(I,x)\psi^-|\am\>\right|} \nonumber\\
& &\!\!\!\!\!\!\!\!\! \!\!\!\!\! \!\!\!\!\!
=
\underset{\sm\in{\mathbb R}}{\rm sup}\,\,
          \underset{n\leq N}{\rm sup}\left(|\sm|+\frac{1}{|\sm|}\right)^{N}
\left|\frac{d^{n}}{d\sm^{n}}\,\,e^{-i\gamma\sqrt{\sm}(t-\bs{x}.\bs{v})}
\<\psi^-|\am\>\right| \nonumber\\
& &\!\!\!\!\!\!\!\!\! \!\!\!\!\! \!\!\!\!\!
=
\underset{\sm\in{\mathbb R}}{\rm sup}\,\,
          \underset{n\leq N}{\rm sup}\left(|\sm|+\frac{1}{|\sm|}\right)^{N}
\left|\sum_{k=0}^{n}\binom{n}{k}
\frac{d^{k}}{d\sm^{k}}e^{-i\gamma\sqrt{\sm}(t-\bs{x}.\bs{v})}
\frac{d^{n-k}}{d\sm^{n-k}}\<\psi^-|\am\>\right|
\label{o2}
\end{eqnarray}
We note that
$$
\frac{d^{k}}{d\sm^k}e^{-i\gamma\sqrt{\sm}(t-\bs{x}.\bs{v})}
=P_{k}\left(\frac{1}{\sqrt{\sm}}\right)e^{-i\gamma\sqrt{\sm}(t-\bs{x}.\bs{v})}
$$
where $P_k(1/\sqrt{\sm})$ is a polynomial in $1/\sqrt{\sm}$.
We also note that $|e^{-i\gamma\sqrt{\sm}(t-\bs{x}.\bs{v})}|\leq 1$
for the chosen branch \eqref{branch} and for $t\geq 0$
(since $\<\psi^-|\am\>\in\tilde{\cal S}\cap\H_{-}^{2}$; and
with $x^{2}\geq0$, $t\geq 0$, we have $t-\bs{x}.\bs{v}\geq 0$).
Thus we obtain for \eqref{o2}
\begin{eqnarray}
\label{o3}
\!\!\!\!\! \!\!\!\!\!
|U(I,x)\psi^{-}|_{N}&\leq&
\underset{\sm\in{\mathbb R}}{\rm sup}\,\,
          \underset{n\leq N}{\rm sup}\left(|\sm|+\frac{1}{|\sm|}\right)^{N}
\sum_{k=0}^{n}\binom{n}{k}\left|P_k\left(\frac{1}{\sqrt{\sm}}\right)\,
\frac{d^{n-k}}{d\sm^{n-k}}\<\psi^-|\am\>\right|\nonumber\\
&\leq&
\underset{n\leq N}{\rm sup}\sum_{k=0}^{n}\binom{n}{k}
\|\Lambda_{n,k}\<\psi^-|\am\>\|_{N}\,,
\end{eqnarray}
where 
$$\Lambda_{n,k}\equiv P_{k}\left(\frac{1}{\sqrt{\sm}}\right)
\frac{d^{n-k}}{d\sm^{n-k}}\,,\qquad n\geq k\,.$$
$\Lambda_{n,k}$ is a $\tau_{\Phi_+}$-continuous operator since
the mappings
$$
\<\psi^{-}|\am\>\mapsto
\frac{d^{n}}{d\sm^{n}}\<\psi^{-}|\am\>\,\,\text{ for }\,\,n=0,1,2,\cdots\,,
$$
and
$$
\<\psi^{-}|\am\>\mapsto
P\left(\frac{1}{\sqrt{\sm}}\right)
\<\psi^{-}|\am\>\,\,\,\text{ for any polynomial }P\,,
$$
are $\tau_{\Phi_+}$-continuous, with Property~(\ref{Property2.2})
justifying the continuity of $P(1/\sqrt{\sm})$.
Continuity of $U(I,x)$ now follows from \eqref{o3}, and the continuity
of $\Lambda_{n,k}$.
Exactly the same arguments apply for $\Phi_-$.

\newpage

\end{document}